\documentclass[12pt]{article}
\usepackage{graphicx}
\usepackage{amssymb}
\newcommand{\req}[1]{(\ref{#1})}
\def\vev#1{\langle #1 \rangle}
\title{Heterotic $F^6$}
\author{S. Stieberger and T.R. Taylor}
\begin{document}
\renewcommand{\thefootnote}{\fnsymbol{footnote}}
\thispagestyle{empty}
\begin{minipage}[t]{5cm}
\small
{\tt hep-th/0207026}\\
July, 2002
\normalsize
\end{minipage}
\hspace{5cm}
\begin{minipage}[t]{4cm}
\begin{flushright}
\small
HU--EP--02/26\\
NUB 3230
\normalsize
\end{flushright}
\end{minipage}

\vspace{1cm}

\begin{center}
{\bf \Large Non-Abelian Born-Infeld Action and \\ Type I --
Heterotic Duality (I):\\[2mm] \large Heterotic
$F^6$ Terms at Two Loops\footnote{Research supported in part by the National
Science Foundation under grant PHY-99-01057, the Deutsche Forschungsgemeinschaft (DFG),
and the German--Israeli Foundation (GIF).}}

\vspace{1.8cm}

S. Stieberger${}^{\,a,b}$ and T.R. Taylor${}^{\,b}$

\vspace{.2cm}

${}^{a}${\em Institut f\"ur Physik, Humboldt Universit\"at
zu Berlin\\
Invalidenstra{\ss}e 110, 10115 Berlin, FRG}

\vspace{.1cm}

${}^{b}${\em Department of Physics, Northeastern University\\ Boston,
MA 02115, USA}

\end{center}

\small\vskip 1cm
\abstract{\noindent
We study the two-loop $F^6$
interactions in $SO(32)$ heterotic superstring
theory in $D{=}10$. By using the generalized Riemann identity we are able to
determine the single-trace part of the
effective action up to a few constants which are related to
certain scattering amplitudes.
This two-loop heterotic result is related by duality
to Type I interactions at the tree level. However, it turns out to be
completely different from any sort of
non-Abelian generalization of Born-Infeld
theory. We offer an explanation of this discrepancy.
}
\noindent

\renewcommand{\thefootnote}{\arabic{footnote}}
\addtocounter{footnote}{-1}
\newpage
\section{Introduction}
Most of the recent progress in superstring theory accomplished by
using various duality conjectures has been limited to the physics
of BPS states. It is very encouraging that certain physical
quantities are completely controlled by short BPS
supermultiplets. Typically, such quantities are highly
constrained (or even uniquely determined) by supersymmetry and
are related to anomalies and topological amplitudes.
An important question arises to what extent duality symmetries
can be tested, or used as a tool,  at the more advanced level of
full-fledged superstring
dynamics involving both short and long supermultiplets.

There is a long-standing problem in Type I superstring
theory which has been waiting for
this type of treatment. It is well known that the classical action for
constant electromagnetic fields associated to
the Cartan subalgebra of $SO(32)$ gauge group generators
is described by Born-Infeld electrodynamics \cite{Tseytlin:1999dj}.
The problem  is to determine the full $SO(32)$ non-Abelian Born-Infeld (NBI)
Lagrangian. The  solution should have the following properties. First,
since the gauge bosons couple to the boundary of a disk, it can be written
as a single trace
of Chan-Paton matrices in the defining representation. In the limit of
weak field strength $F,$ this would give  an expansion of
the form Tr$\sum_{n=0}^{\infty}b_nF^{2n}$ with some constant coefficients
$b_n$. Finally, when $F$'s are restricted to the Cartan subalgebra,
one should recover the Born-Infeld Lagrangian. Soon after Type I -- heterotic
duality was conjectured \cite{Polchinski:1995df, apt}, Tseytlin
\cite{Tseytlin:1995fy} pointed out
that Type I tree-level $F^{2n}$ interactions appear
on the heterotic side at genus $g=n{-}1$. For instance, the classical
$F^4$ interactions of open superstrings \cite{Tseytlin:1986ti}
are related to the one-loop elliptic genus of the heterotic superstring
\cite{Lerche:1987sg,Ellis:1987dc}. With some little optimism, one could
try to go to higher genus in order to obtain further terms in the $F^{2n}$
expansion of the non-Abelian Born-Infeld action. The reason why
some optimism is needed is that Type I -- heterotic correspondence
is a strong-weak coupling
duality so it is guaranteed to work in a straightforward manner
only for ``BPS-saturated'' quantities that
are subject to non-renormalization theorems. An order by order comparison
of generic quantities can be more subtle, nevertheless it is more
interesting because renormalized quantities are usually not as much restricted
by (super)symmetries as the non-renormalized ones.

Keeping all these reservations in mind, we present in this paper an analysis of
$F^6$ interaction terms in the heterotic superstring theory at two loops
($g{=}2\leftrightarrow n{=}3$). The problem of renormalization
will be discussed in a separate publication \cite{part1}.
Our goal is to extract the effective action from the six-point
scattering amplitudes. The paper is organized as follows. In Section 2, we
outline the strategy and explain some technical points.
In Section 3, we study
the simpler case of $F^{4}$ terms and show how the Tr$(t_8F^{4})$
structure arises as a consequence of Riemann identity. In Section 4,
we discuss the $F^6$ terms for the field strengths associated to the
Cartan subalgebra of $SO(32)$.
The result turns out to be different from the expression obtained by
expanding the Abelian Born-Infeld action. In Section 5, we extend
our results to the full SO(32) field strength. We parameterize
the two-loop $F^{6}$ Lagrangian in terms of a few constants related to certain
scattering amplitudes. In Section 6 we recapitulate the results
in the context of Type I -- heterotic duality.

\section{Generalities}
At first sight, two-loop computations of heterotic
scattering amplitudes seem to be more challenging than analogous
tree-level computations in Type I superstring theory. There is however,
an important simplification which is due to the form of gauge field
vertex operators. In the 0-ghost picture \cite{Friedan:1985ge},
the heterotic vertex operator for a gauge
boson of momentum $p$ and polarization $\epsilon$ is:
\begin{equation}
V(\epsilon, p;z,\bar{z})=\; :\!(\epsilon\!\cdot\!\partial X+
ip\!\cdot\!\psi\,\epsilon\!\cdot\!\psi)
\bar{J}^ae^{ip\cdot X}\!\!:\; ,\label{vertex}
\end{equation}
where $\psi$'s are the spacetime fermions and
$\bar{J}^a$ is the left-moving Ka\v{c}-Moody current. At genus one and higher,
summations over spin structures always require some number of $\psi$
insertions in order to avoid cancellations due to Riemann identities
in the CP even sectors or to
saturate zero modes in the CP odd sectors. These insertions are provided by the
fermionic part of (\ref{vertex}) which involves a momentum factor and
creates directly the
gauge field strength $F_{\mu\nu}$. This  facilitates the task of identifying the
corresponding terms in the low-energy effective action, in
comparison to Type I theory at the tree level, where
the momentum dependence arises also from the bosonic contractions involving
the $e^{ip\cdot X}$ factors that yield hypergeometric functions
\cite{Tseytlin:1986ti} and often
require momentum expansions and elimination of reducible contributions
before constructing a gauge invariant low-energy action. For amplitudes
involving five or more gauge bosons it becomes a very laborious procedure.

In this paper, we focus on the CP even part of scattering amplitudes.
We  analyze the kinematic configurations that lead to amplitudes amenable to
simplification by using the two-loop bosonisation formulae \cite{Verlinde:1986kw}
or equivalently,
the Fay trisecant identity \cite{Fay,Mumford}. We identify a set of amplitudes that
vanish as a consequence of generalized Riemann identity \cite{Fay,Mumford},
that is as a result of
spacetime supersymmetry.
On the effective field theory side,
we make an ansatz for the non-Abelian Born-Infeld
action as a series consisting of Tr$F^{2n}$ terms with
a single $SO(32)$ gauge trace Tr in the defining representation
and with all possible $SO(10)$ Lorentz group contractions.
Such an action contains, for instance, 4 inequivalent $F^4$ terms,
31 inequivalent $F^6$ terms etc.
By comparing the vanishing amplitudes with this form of NBI action,
we derive certain relations between the coefficients of various terms
and parameterize
the solution in terms of a smaller number of constants. These constants can be in
principle calculated from superstring theory, although in practice it
may be quite a difficult task since as we shall see later on,
beyond the $F^4$ order,
the relevant amplitudes couple non-trivially the left- and right-moving sectors,
therefore they do not seem to have a topological character.

When the gauge boson vertex operators are inserted on a genus 2 Riemann
surface in the 0-ghost picture (\ref{vertex}), the amplitude
involves also 2 picture-changing
operators (PCOs)
\begin{equation}
Y=-{1\over 2}e^{\phi}\psi\cdot\partial X+ \makebox{ghost terms,}\label{pco}
\end{equation}
where $\phi$ is the free scalar bosonizing the superghost system
\cite{Friedan:1985ge}.
These operators can be inserted at arbitrary points $x_1$ and $x_2$.
Thus we will study the following correlation functions:
\begin{equation}\label{smatrix}
\vev{Y(x_1)Y(x_2)\prod_{i=1}^{2n} V(\epsilon_i,p_i;z_i,\bar{z}_i)}\ .
\end{equation}
We will first focus on the contributions involving the ``matter'' part of
PCOs, i.e.\ the term shown explicitly in Eq.(\ref{pco}) and the fermionic part of
gauge boson vertices (\ref{vertex}). Since we are interested in $F^{2n}$ terms
and now the vertices  bring at least
one power of momentum for each gauge boson, the momenta of the
exponentials in \req{vertex} can be set to
zero.\footnote{In principle, this step needs some care, because the integration
over the vertex positions $z_i$ may lead to singularities from colliding points.
These pinching effects, however, do not affect our discussion.}
Then the right-moving (superstring) part of these
contributions acquires
a generic form:
\begin{equation}
\dots\sum_{\delta\,\makebox{\scriptsize even}}
\langle e^{\phi(x_1)}e^{\phi(x_2)}\rangle_{\delta}\,\langle
\psi(x_1)\!\cdot\!\psi(x_2)
\prod_{i=1}^{2n}p_i\!\cdot\!\psi(z_i)\,
\epsilon_i\!\cdot\!\psi(z_i)\rangle_{\delta}\, ,\label{amp}
\end{equation}
where the sum extends over even spin structures $\delta$ and the ellipsis
represents some factors that do not depend  on spin structures.
All these correlators involve Riemann $\theta$-functions.
Before discussing them in more detail, let us recall
that Riemann identity  can be written in the following form \cite{Mumford}:
\begin{equation}
\sum_{\delta}\langle\alpha|\delta\rangle\,\theta_\delta(z_1)
\theta_\delta(z_2)\theta_\delta(z_3)\theta_\delta(z_4)=2^g\,\theta_\alpha(z_1')
\theta_\alpha(z_2')\theta_\alpha(z_3')\theta_\alpha(z_4')\, ,\label{riemann}
\end{equation}
where $\theta_{\delta}$ denotes the genus $g$ $\theta$-function of spin structure
$\delta$, $\alpha$ is an arbitrary (reference) spin structure and the phase factor
\begin{equation}
\langle\alpha|\delta\rangle\equiv e^{4\pi i(\vec{\delta}_1\cdot\vec{\alpha}_2
-\vec{\delta}_2\cdot\vec{\alpha}_1)}=\pm 1\, .
\end{equation}
The transformation of arguments is given by
\begin{equation}\left(\!\begin{array}{c}z_1'\\ z_2'\\ z_3'\\ z_4'
\end{array}\!\right)={1\over 2}\left(\!\begin{array}{rrrr} 1&1&1&1\\ 1&1&-1&-1\\
1&-1&1&-1\\ 1&-1&-1&1\end{array}\!\right)\!\!\left(\!\begin{array}{c}z_1\\ z_2\\
z_3\\ z_4 \end{array}\!\right).
\end{equation}
Note that the sum is over {\em all\/} spin structures, at $g{=}2$ 10 even and 6
odd ones.

There is a specific choice of the PCO insertion points that is particularly
suitable for applying the Riemann identity (\ref{riemann}).
In the so-called unitary gauge
\cite{Verlinde:1987sd,Lechtenfeld:1989ke,MP},
the points $x_1$ and $x_2$ are chosen such that
\begin{equation}
x_1+x_2=2\Delta_{\alpha}\, ,\label{choice}
\end{equation}
where $\Delta_{\alpha}$ is the Riemann $\theta$-constant which represents the degree
1 divisor of half-differentials associated with the
spin structure $\alpha$ \cite{Mumford}. In this case,
\begin{equation}
\langle e^{\phi(x_1)}e^{\phi(x_2)}\rangle_{\delta}~\propto~
\langle\alpha|\delta\rangle\,\theta_{\delta}(0)^{-1},\label{phis}
\end{equation}
and, as we shall see below, for certain kinematic
configurations, the superstring amplitudes (\ref{amp})
look exactly like the l.h.s.\ of Eq.(\ref{riemann}).

\section{$F^4$ from 4-point amplitudes}
The heterotic $F^4$ terms have been discussed before in many places,
at one-loop \cite{Ellis:1987dc}
and, to certain extent, at two-loop level \cite{iengo,ellistwo,Lechtenfeld:1989ke},
however it is instructive to discuss them here as a preparation for
studying $F^6$. The NBI ansatz has the form:
\begin{eqnarray}
{\cal L}_4 &=& a_1{F_{\alpha_1\alpha_2}}
{F_{\alpha_2\alpha_3}}
{F_{\alpha_3\alpha_4}}{F_{\alpha_4\alpha_1}} + a_2{F_{\alpha_1\alpha_2}}
{F_{\alpha_3\alpha_4}}{F_{\alpha_2\alpha_3}}
{F_{\alpha_4\alpha_1}}\nonumber\\
&&\hskip -3.5mm + a_3 \,{F_{\alpha_1\alpha_2}}
{F_{\beta_1\beta_2}}{F_{\alpha_2\alpha_1}}
{F_{\beta_2\beta_1}}\, +a_4\,{F_{\alpha_1\alpha_2}}{F_{\alpha_2\alpha_1}}
{F_{\beta_1\beta_2}}
{F_{\beta_2\beta_1}}\, ,\label{f4}
\end{eqnarray}
with the understanding that the $SO(32)$ gauge field strength matrices
are taken in the defining representation and one, overall $SO(32)$ trace
Tr is taken of the whole expression. It is very convenient to display
these contributions diagrammatically, by drawing points on the corners
of a square
indicating the position of the $F$-factors in the trace, and with the solid
lines representing {\em Lorentz\/} contractions \cite{Sevrin:2001ha}, see Fig.1.
Thus $a_1$ is the coefficient of the square diagram, $a_2$
is the bow tie, $a_3$ the cross, and $a_4$
the two parallel lines. It is important to keep in mind that as far as the gauge
group indices are concerned, all these diagrams
represent {\em one\/} type of contraction which, in the
same spirit, could be represented as a square.
\begin{figure}[h]\hfill
\includegraphics[scale=.2]{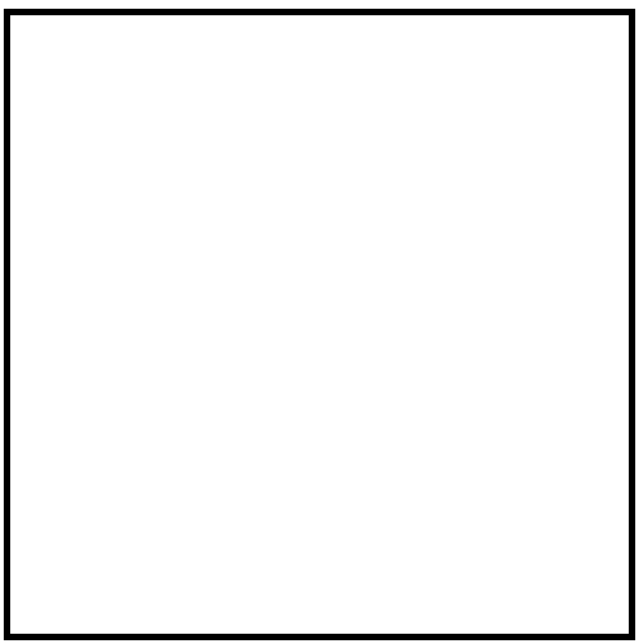}\hskip 1cm
\includegraphics[scale=.2]{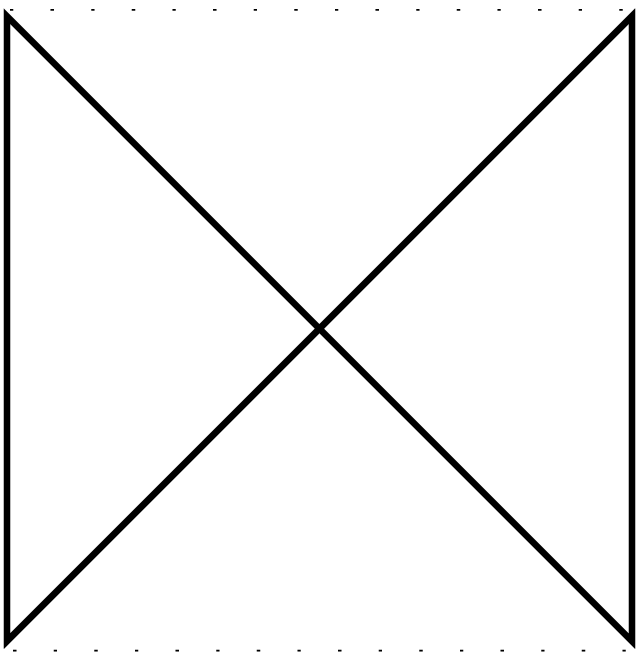}\hskip 1cm
\includegraphics[scale=.2]{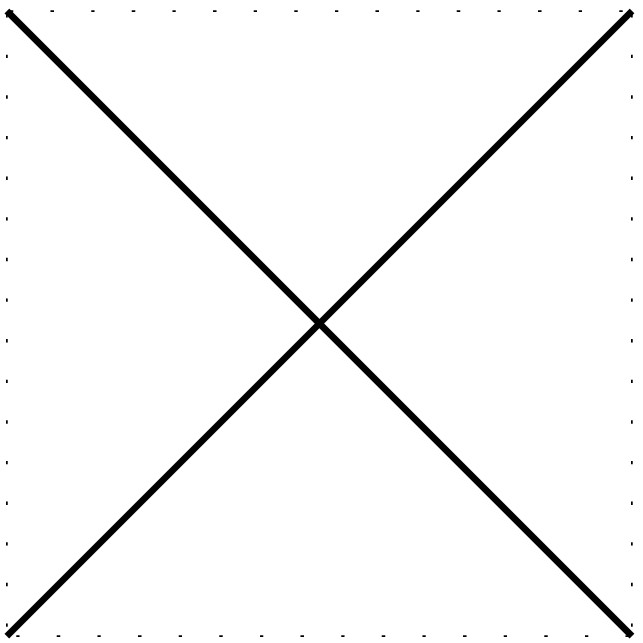}\hskip 1cm
\includegraphics[scale=.2]{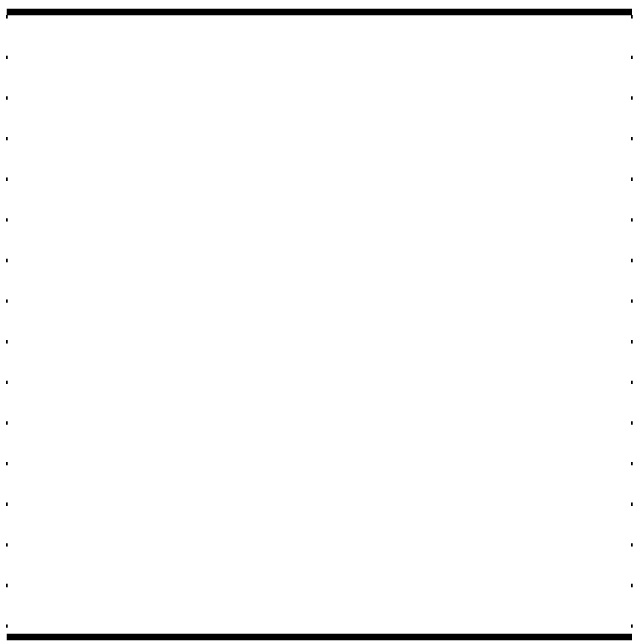}\hfill\hfill
\caption{Diagrammatic  representation of the four terms in Eq.(\ref{f4}).}
\label{f4fig}
\end{figure}

In the heterotic amplitudes, the structure of gauge group factors
is governed by the correlator of left-moving Ka\v{c}-Moody currents.
Without going into detail, we can focus on one
particular contribution with the square contraction of {\em gauge\/}
indices.

The dependence of the amplitude (\ref{amp}) on spacetime momenta
and polarizations is governed by
\begin{equation} {\cal Z}_{\delta}(x,z;p,\epsilon)\equiv
\langle\alpha|\delta\rangle\,\theta_{\delta}(0)^{-1}\langle
\psi(x_1)\!\cdot\!\psi(x_2)
\prod_{i=1}^{4}p_i\!\cdot\!\psi(z_i)\,
\epsilon_i\!\cdot\!\psi(z_i)\rangle_{\delta}\, .\label{zdelta}
\end{equation}
Note the prefactor coming from two-point function (\ref{phis});
{\em all\/} spin structure dependent contributions are now contained
in ${\cal Z}_{\delta}$. For an arbitrary kinematic
configuration, the fermionic correlator generates a large number
of terms, with the scalar products of polarization vectors and
momenta dictated by Wick contractions.
{}For our purposes, it is convenient to split $D{=}10$ spacetime into
five complex planes so that within each plane
\begin{equation}
p\cdot\psi=p^-\psi+p^+\bar{\psi}\, .\label{prod}
\end{equation}
The two-point fermion correlation function is given by
\cite{Verlinde:1986kw}\footnote{This
function is defined as the Polyakov path integral, without
the $\theta_{\delta}(0)^{-1}$ normalization factor, which is different from the
conventions used in \cite{Lechtenfeld:1989ke}.}
\begin{equation}
Z_{\delta}(z,w)=\langle
\bar{\psi}(z)\psi(w) \rangle_{\delta}=\theta_{\delta}(z-w)\,
E(z,w)^{-1}\, ,
\end{equation}
where $E(z,w)$ is the prime form (antisymmetric in $z\leftrightarrow w$)
\cite{Fay}.
A generic contribution to ${\cal Z}_{\delta}$
involves 5 position-dependent theta function factors -
too many for a direct application of Riemann's identity (\ref{riemann}).
One way to reduce the number of position-dependent theta functions is to use
Fay's trisecant identity \cite{Fay, Mumford}:
\begin{equation}
Z_{\delta}(z_1,w_1)Z_{\delta}(z_2,w_2)-
Z_{\delta}(z_2,w_1)Z_{\delta}(z_1,w_2)=\theta_{\delta}(0)\,
Z_{\delta}(z_1,z_2,w_1,w_2)\, ,\label{fay}
\end{equation}
where, for 2 and more (distinct) points,  $Z_{\delta}$
is defined as the chiral determinant amplitude \cite{Verlinde:1986kw}
\begin{eqnarray}
Z_{\delta}(z_1,\dots,z_m,w_1,\dots,
w_m)&\equiv&\langle
\bar{\psi}(z_1)\dots\bar{\psi}(z_m)\psi(w_1)\dots \psi(w_m)
\rangle_{\delta}\nonumber\\[3mm]
&&\hskip -4.2cm =\, \theta_{\delta}(0)^{1{-}m}
\makebox{det}_{ij}\left[ {\theta_{\delta}( z_i- w_j)
\over E(z_i,w_j)}\right]\label{chiral}\\[3mm]
=\,\theta_{\delta}(\textstyle\sum z_i-\sum w_j)&&\hskip -1cm
{\displaystyle \prod_{i<j}E(z_i,z_j)\prod_{i<j}E(w_i,w_j)\over\displaystyle
\prod_{i,j}E(z_i,w_j)}.\nonumber
\end{eqnarray}
Here, we will proceed in a slightly different way, by selecting kinematic
configurations that lead directly to chiral determinants of
Eq.(\ref{chiral}), therefore contain a smaller number
of position-dependent theta functions from the outset.
This is mathematically equivalent to using Fay's identity, but
is more convenient for comparing with the low-energy
Lagrangian (\ref{f4}).

Consider the momenta and polarization vectors of the
four bosons distributed on
3 complex planes as follows:
\begin{equation} \begin{tabular}{c}
$p_1~ p_2~ p_3~ p_4$\\ \hline\\[-2mm]
$\epsilon_1~ \epsilon_2$ \\ \hline\\[-2mm]
$\epsilon_3~ \epsilon_4$\\
\hline
\end{tabular}\end{equation}
Due to orthogonality, only a limited number of
fermion contractions contribute now to the
amplitude (\ref{zdelta}). Note that the two PCO
fermions can be contracted between themselves or with the
fermions in one of the planes.
These two types of PCO contractions are called ``split'' and ``non-split,''
respectively \cite{Lechtenfeld:1989ke}. We first discuss the non-split case and,
to be specific, we pick up
the part of $\psi(x_1)\!\cdot\!\psi(x_2)$ from the plane associated
to $(\epsilon_3, \epsilon_4)$, so that the fermion correlator factorizes as
\begin{equation}
\langle\alpha|\delta\rangle\,\theta_{\delta}(0)^{-1}\langle
\prod_{i=1}^{4}p_i\!\cdot\!\psi(z_i)\rangle\langle
\epsilon_1\!\cdot\!\psi(z_1)\epsilon_2\!\cdot\!\psi(z_2)\rangle
\langle\psi(x_1)\!\cdot\!\psi(x_2)\epsilon_3\!\cdot\!
\psi(z_3)\epsilon_4\!\cdot\!\psi(z_4)\rangle\label{nonsplit}
\end{equation}
Once the scalar products are written in the complex basis,
c.f.\ Eq.(\ref{prod}), each plane contributes a combination of
chiral determinants (\ref{chiral}). Let $[(p_1^{\pm}p_2^{\pm}
p_3^{\pm}
p_4^{\pm})(\epsilon_1^{\pm}\epsilon_2^{\pm})(\epsilon_3^{\pm}
\epsilon_4^{\pm})]$ denote coefficients of the respective kinematic factors.
We are particularly interested in
\begin{eqnarray}
[(p_1^+p_2^-p_3^+p_4^-)(\epsilon_1^-\epsilon_2^+)(\epsilon_3^-
\epsilon_4^+)]_{\makebox{\scriptsize non-split}}
&=& \nonumber\\[2mm] &&\hskip -6cm
\dots\sum_{\delta\,\makebox{\scriptsize even}}
\langle\alpha|\delta\rangle\,\theta_{\delta}(0)
\theta_{\delta}(z_1{-}z_2{+}z_3{-}z_4)
\theta_{\delta}(z_2{-}z_1)\theta_{\delta}(x_1{-}x_2{+}z_4{-}z_3)\nonumber \\
&&\hskip -4.5cm  +~(x_1\leftrightarrow x_2)\, .
\end{eqnarray}
Note that since the two ``empty'' planes bring two additional
determinants $\theta_{\delta}(0)$, we are left with just one
$\theta_{\delta}(0)$ factor. Although the sum is
over the even spin structures only, it can be formally extended to all spin
structures because $\theta_{\delta}(0)=0$ for  odd $\delta$. At this point,
we can apply Riemann's identity (\ref{riemann}) to obtain
\begin{eqnarray}
[(p_1^+p_2^-p_3^+p_4^-)(\epsilon_1^-\epsilon_2^+)(\epsilon_3^-
\epsilon_4^+)]_{\makebox{\scriptsize non-split}}&=&  \label{summa}\\[2mm]
&&\hskip -6.5cm
\theta_{\alpha}(\Delta_{\alpha}{-}x_2) \theta_{\alpha}(\Delta_{\alpha}{-}
x_1{+}z_1{-}z_2{+}z_3{-}z_4)\nonumber
\theta_{\alpha}(\Delta_{\alpha}{-}x_1{+}z_2{-}z_1)
\theta_{\alpha}(\Delta_{\alpha}{-}x_2{+}z_4{-}z_3)\, ,
\end{eqnarray}
where we used the relation (\ref{choice}). Since
\begin{equation}\theta_{\alpha}(\Delta_{\alpha}-z)=0 \label{zero}\label{del}
\end{equation}
for any point $z$ on the Riemann surface, the first factor on the r.h.s.\ of
(\ref{summa})
gives zero, therefore the PCO contractions with fermions associated to the
$(\epsilon_3,\epsilon_4)$ plane
do not contribute to the kinematic factor under considerations.
Similar arguments apply to
$(p_1,p_2,p_3,p_4)$ and $(\epsilon_1,\epsilon_2)$ planes.

We now turn to the split contributions of PCO fermions
associated to two orthogonal (``empty'') planes. The two fermions at $x_1$ and $x_2$
are now contracted with each other and we obtain
\begin{eqnarray}
[(p_1^+p_2^-p_3^+p_4^-)(\epsilon_1^-\epsilon_2^+)(\epsilon_3^-
\epsilon_4^+)]_{\makebox{\scriptsize split}}&=& \nonumber\\[2mm] &&\hskip -6cm
\dots\sum_{\delta\,\makebox{\scriptsize even}}
\langle\alpha|\delta\rangle\,\theta_{\delta}(x_1{-}x_2)
\theta_{\delta}(z_1{-}z_2{+}z_3{-}z_4)
\theta_{\delta}(z_2{-}z_1)\theta_{\delta}(z_4{-}z_3)\nonumber \\
&&\hskip -4.5cm  +~(x_1\leftrightarrow x_2)\, .
\end{eqnarray}
The Riemann identity, applied to the sum over {\em all\/}
spin structures, gives
\begin{eqnarray}
\sum_{\delta}
\langle\alpha|\delta\rangle\,\theta_{\delta}(x_1{-}x_2)
\theta_{\delta}(z_1{-}z_2{+}z_3{-}z_4)
\theta_{\delta}(z_2{-}z_1)\theta_{\delta}(z_4{-}z_3)&=&\nonumber\\
&& \hskip -10.3cm
\theta_{\alpha}(\Delta_{\alpha}{-}x_2) \theta_{\alpha}(\Delta_{\alpha}{-}
x_2{+}z_1{-}z_2{+}z_3{-}z_4)
\theta_{\alpha}(\Delta_{\alpha}{-}x_2{+}z_2{-}z_1)
\theta_{\alpha}(\Delta_{\alpha}{-}x_2{+}z_4{-}z_3)\nonumber\\[1mm]&=&
0\label{all}\, ,
\end{eqnarray}
where we used Eq.(\ref{del}) again. Interchanging $x_1$ and $x_2$ one
finds similarly a zero result therefore the sums over
even and odd spin structures,
which correspond to the l.h.s.\ of Eq.(\ref{all})
symmetrized and antisymmetrized in  $(x_1,x_2)$, respectively, must
vanish
separately. Hence for the present
kinematic factor, the split PCO fermion contributions
give zero result.
Note that, in principle, the ghost parts of PCOs could give rise
to non-vanishing self-contractions however, as shown by Lechtenfeld
and Parkes \cite{Lechtenfeld:1989ke},
the structure of such terms is very similar to
those of split fermions so if the latter are zero by Riemann
identity, the ghost contractions cancel as well.
Thus
\begin{equation}
[(p_1^+p_2^-p_3^+p_4^-)(\epsilon_1^-\epsilon_2^+)(\epsilon_3^-
\epsilon_4^+)]~=~0\, .\label{final}
\end{equation}

So far our discussion has been limited to the fermionic parts
of gauge boson
vertices (\ref{vertex}). However, if one or more fermion
bilinears are replaced
by bosons,  the correlators involve a smaller number of spin
structure-dependent
theta functions and all
arguments based on the Riemann identity remain valid.
In particular,
Eq.(\ref{final}) holds for the {\em full\/} superstring
amplitude.
{}Furthermore, it is obvious
that all kinematic factors obtained from the present one
by permuting the gauge boson indices (1,2,3,4) are also vanishing.
We will show that this information is completely sufficient for
determining the coefficients of NBI ansatz (\ref{f4}) modulo
one multiplicative constant.

In order to obtain the scattering amplitudes from the NBI ansatz
(\ref{f4}), we apply the following procedure. First, inside
each of $F^4$ terms, we label $F$'s by the indices $j=1,2,3,4$,  for
instance
\begin{equation} {F_{\alpha_1\alpha_2}}
{F_{\alpha_2\alpha_3}}
{F_{\alpha_3\alpha_4}}{F_{\alpha_4\alpha_1}}\rightarrow
{F_{1\alpha_1\alpha_2}} {F_{2\alpha_2\alpha_3}}
{F_{3\alpha_3\alpha_4}}{F_{4\alpha_4\alpha_1}}
\end{equation}
Next, in Eq.(\ref{f4}) we substitute
\begin{equation}
F_{j\alpha\beta}\rightarrow
(p_{j\alpha}\epsilon_{j\beta}-p_{j\beta}\epsilon_{j\alpha})
\lambda_{j}\, ,
\end{equation}
where $\lambda_j$ is the $SO(32)$ charge of
$j^{\makebox{\scriptsize th}}$ gauge boson (a matrix in the
defining representation). Finally, we symmetrize in the indices
(1,2,3,4). Note that the symmetrization gives rise to gauge group
traces ordered in all possible ways, so one particular gauge group
structure, like the square contraction, accompanies a kinematic
factor corresponding to a cyclic orbit. After rewriting the scalar
products in the complex basis, we can extract the relevant
kinematic coefficients and compare with Eq.(\ref{final}) and its
permutations. In this way, we obtain one equation per each
permutation, however only three of them are linearly independent:
\begin{eqnarray}
a_1+2a_4&=& 0\, \nonumber\\
a_2+4a_4&=& 0\, \label{as}\\
a_2+8a_3&=& 0\, \, .\nonumber
\end{eqnarray}
Their solution can be written as
\begin{equation}
a_1=a\qquad a_2=2a\qquad a_3=-\frac{a}{4} \qquad
a_4=-\frac{a}{2}\, ,\label{bisym}
\end{equation}
where $a$ is an arbitrary constant. The corresponding combination of
$F^4$ terms is the well-known $t_8$ contraction that appears already
at the one loop level \cite{Lerche:1987sg,Ellis:1987dc}.
As emphasized by Tseytlin
\cite{Tseytlin:1997cs}, this combination
can be obtained by the ``symmetric trace'' prescription from
the Abelian Born-Infeld action.
At one loop, the coefficient $a=\frac{1}{24}$ is related to Green-Schwarz anomaly.
The fact that the $t_8$ combination reappears at higher loops
has been known already for some time \cite{iengo,ellistwo,Lechtenfeld:1989ke}
but only recently \cite{part1},
we have explicitly shown that $a=0$ at two loops.
The derivation of relations (\ref{as}) from Riemann identity
has been presented here as a warm-up for the more complicated $F^6$ case.
\section{Abelian $F^6$}
A complete discussion of $F^6$ terms will be given in the next section.
Since the bottom line will turn out to be quite surprising, we chose
to first discuss  the field strengths associated to
$SO(32)$ Cartan subalgebra generators. The most general Abelian
action for one of these generators reads:
\begin{eqnarray}
{\cal L}_{6A} &=& h\,{F_{\alpha_1\alpha_2}}
{F_{\alpha_2\alpha_3}}
{F_{\alpha_3\alpha_4}}{F_{\alpha_4\alpha_5}} {F_{\alpha_5\alpha_6}}
{F_{\alpha_6\alpha_1}} \nonumber\\ &&\hskip -3.5mm
+\,s\,{F_{\beta_1\beta_2}}{F_{\beta_2\beta_1}}{F_{\alpha_1\alpha_2}}
{F_{\alpha_2\alpha_3}}{F_{\alpha_3\alpha_4}}
{F_{\alpha_4\alpha_1}}
\nonumber\\
&&\hskip -3mm +\, l\,{F_{\alpha_1\alpha_2}}
{F_{\alpha_2\alpha_1}}{F_{\beta_1\beta_2}}
{F_{\gamma_1\gamma_2}}{F_{\gamma_2\gamma_1}{F_{\beta_2\beta_1}}}.
\label{f6ab}
\end{eqnarray}
The diagrams representing these terms are shown
in Fig.2. Thus $h$ is associated to the hexagon, $s$ is the square contraction
accompanied by a line and $l$ is associated to three lines.
\begin{figure}[h]\hfill
\includegraphics[scale=.2]{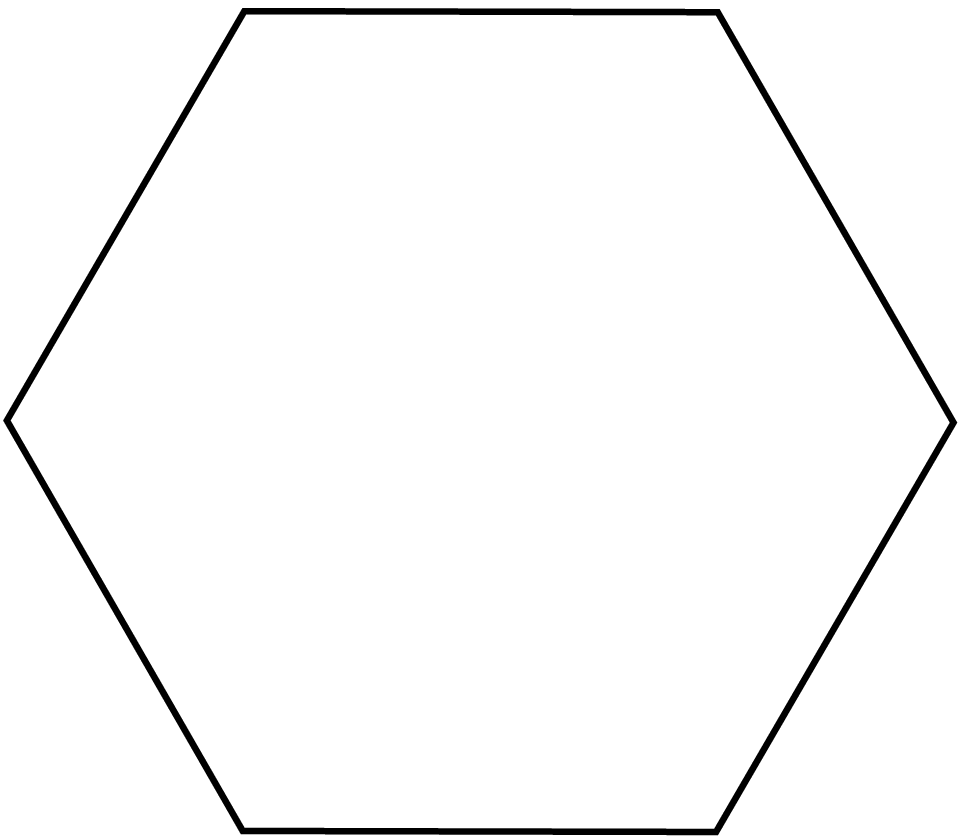}\hskip 1.5cm
\includegraphics[scale=.2]{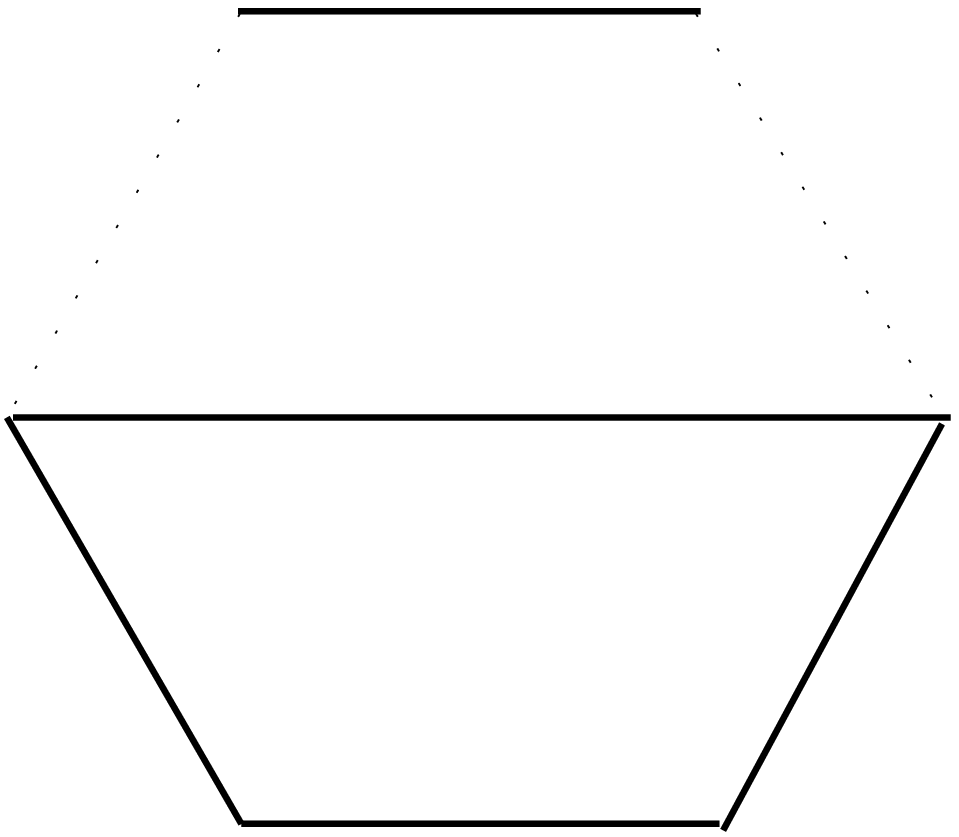}\hskip 1.5cm
\includegraphics[scale=.2]{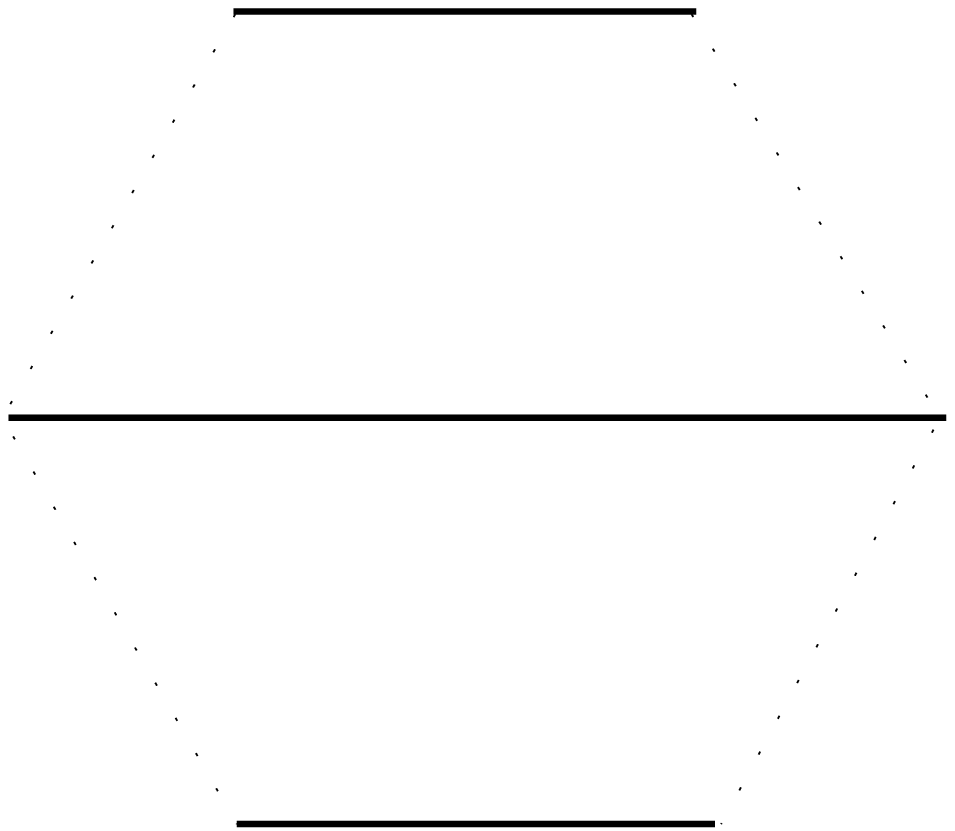}\hfill\hfill
\caption{Diagrammatic  representation of the three terms in Eq.(\ref{f6ab}).}
\label{f6abfig}
\end{figure}

{}For the commuting generators, the correlator of Ka\v{c}-Moody currents
is completely symmetric in the vertex positions $z_i$.
{}From the superstring side of gauge boson vertices, we first pick up
the purely fermionic parts and focus on ``split'' contributions.
The fermion contractions give rise to
three types of terms:
\begin{eqnarray}H&=&\int\prod_{i=1}^6 d^2z_i\dots \label{hhh}
\sum_{\delta\,\makebox{\scriptsize even}}
\langle\alpha|\delta\rangle\,\theta_{\delta}(0)^{-3}\,\theta_{\delta}(x_1-x_2)
\times
\nonumber\\ &&\hskip -1cm
\theta_{\delta}(z_1{-}z_2)\theta_{\delta}(z_2{-}z_3)\theta_{\delta}(z_3{-}z_4)
\theta_{\delta}(z_4{-}z_5)\theta_{\delta}(z_5{-}z_6)\theta_{\delta}(z_6{-}z_1)\\
S&=&\int\prod_{i=1}^6 d^2z_i\dots \label{sss}
\sum_{\delta\,\makebox{\scriptsize even}}
\langle\alpha|\delta\rangle\,\theta_{\delta}(0)^{-3}\,\theta_{\delta}(x_1-x_2)
\times
\nonumber\\ &&\hskip -1cm
\theta_{\delta}(z_1{-}z_2)\theta_{\delta}(z_2{-}z_1)\theta_{\delta}(z_3{-}z_4)
\theta_{\delta}(z_4{-}z_5)\theta_{\delta}(z_5{-}z_6)\theta_{\delta}(z_6{-}z_3)\\
L&=&\int\prod_{i=1}^6 d^2z_i\dots\label{ccc}
\sum_{\delta\,\makebox{\scriptsize even}}
\langle\alpha|\delta\rangle\,\theta_{\delta}(0)^{-3}\,\theta_{\delta}(x_1-x_2)
\times
\nonumber\\ &&\hskip -1cm
\theta_{\delta}(z_1{-}z_2)\theta_{\delta}(z_2{-}z_1)\theta_{\delta}(z_3{-}z_4)
\theta_{\delta}(z_4{-}z_3)\theta_{\delta}(z_5{-}z_6)\theta_{\delta}(z_6{-}z_5)
\end{eqnarray}
The above expressions can be represented diagrammatically as on Fig.2,
but now with theta functions instead of $F$'s and the
arguments $z_i{-}z_j$ playing the role of Lorentz indices. We will show
below that
\begin{equation}
H=0\; ,\quad L=-3S\, .\label{nonbi}
\end{equation}

In order to prove Eqs.(\ref{nonbi}), we start from Riemann identity applied
to two sets of theta function arguments:
\begin{eqnarray} &&\hskip -1cm
\sum_{\delta\,\makebox{\scriptsize even}}
\langle\alpha|\delta\rangle\,\theta_{\delta}(x_1{-}x_2)\times\nonumber\\&&
\theta_{\delta}(z_1{+}z_5{-}z_2{-}z_6)
\theta_{\delta}(z_2{+}z_4{-}z_3{-}z_5)\theta_{\delta}(z_3{+}z_6{-}z_1{-}z_4)
=0\, ,\\[3mm] &&\hskip -1cm
\sum_{\delta\,\makebox{\scriptsize even}}
\langle\alpha|\delta\rangle\,\theta_{\delta}(x_1{-}x_2)\times\nonumber\\&&
\theta_{\delta}(z_1{+}z_2{+}z_3{-}z_4{-}z_5{-}z_6)
\theta_{\delta}(z_4{-}z_3)\theta_{\delta}(z_5{+}z_6{-}z_1{-}z_2)=0\, .
\end{eqnarray}
Both equations follow from Eq.(\ref{riemann}) after using Eq.(\ref{del})
and symmetrization in $x_1, x_2$. Theta functions of
arguments involving more than two points can be expressed in terms of
the basic two-point functions by using the
chiral determinant formula (\ref{chiral}), giving rise to the products
appearing in
the integrands of Eqs.(\ref{hhh}-\ref{ccc}), modulo some permutations of
the integration variables.
Taking into account the symmetry of full integrands under
these permutations, we obtain:
\begin{eqnarray}\hskip -.7cm
&&3H+3S+L=0\, ,\label{33l}\\ \hskip -.7cm
&&4H+6S+2L=0\label{46l}\, .
\end{eqnarray}
Eq.(\ref{nonbi}) is the solution of the above equations.

Since the Lorentz contractions in kinematic factors arise from
the world-sheet fermion contractions, the integrals of Eqs.(\ref{hhh}-\ref{ccc})
are directly related to the coefficients of the effective action (\ref{f6ab}).
It is a matter of simple combinatorics to show that
\begin{equation}
h={H\over 12}\, ,\qquad s={S\over 32}\, ,\qquad l={L\over 384}\, .\label{comb}
\end{equation}
Although the proof Eq.(\ref{nonbi}) has been given for the ``split" terms
only,
similar arguments apply to all ``non-split'' contributions to the three
types of kinematic factors.
Hence Eq.(\ref{nonbi}) is valid for the full string amplitude and
\begin{equation}
h=0\, , \qquad s=-4 l\, .\label{hsl}
\end{equation}
The above result, which will be rederived in the next Section as
a limit of the full non-Abelian action, is quite surprising. The
low-energy action turns out to be different from the one obtained
by expanding the Born-Infeld Lagrangian \cite{Tseytlin:1999dj}:
$h={1\over 12}$, $s=-{1\over 32}$, and $l={1\over 384}$ (or
$H{=}L{=}{-}S{=}1$).\nopagebreak

It has been previously argued in \cite{MT} that the only form of
$F^6$ terms allowed in $D{=}10$ by $N{=}1$ supersymmetry is the
above Born-Infeld combination.\footnote{We are grateful to Arkady
Tseytlin for bringing this problem to our attention.} The result
(\ref{hsl}) seems to contradict this claim unless the two-loop
coefficient vanishes exactly after combining all contributions,
which is obviously consistent with supersymmetry because $F^4$
together with the related fermionic terms form a complete
superinvariant. However, the arguments of \cite{MT} rely on one
specific form of supersymmetry transformations, therefore a
careful examination of their possible deformations is required
before jumping to the conclusion that this is indeed the case.
Note that when trivially compactified to $D{=}4$, Eq.(\ref{hsl})
agrees with the supersymmetric Born-Infeld Lagrangian \cite{CF}:
the Lorentz hexagon degenerates in $D{=}4$. In either case, we
find that already in the Abelian limit, the heterotic two-loop
result does not match the tree-level Type I theory. We will
discuss this discrepancy later.

\section{Non-Abelian $F^6$}
The most general, single-trace non-Abelian $F^6$ action contains
31 terms with various Lorentz contractions and orderings of
$F$'s. 28 of them can be generated from the diagrams shown in
Fig.2 in the following way \cite{Sevrin:2001ha}. First, the
corners are enumerated by 1 to 6 in the clockwise order, starting
from the north-west corner. Then the corners are permuted in
certain ways, producing new diagrams. Instead of drawing these
diagrams (they can be found in Ref.\cite{Sevrin:2001ha}), we
prefer to list the relevant permutations in Table 1. Their
coefficients  will be denoted by $h(i)$, $i=1,\dots, 14$, $s(j)$,
$j=1,\dots, 9$, and and $l(k)$, $k=1,\dots,5$, for the diagrams
generated by the hexagon, square plus one line, and three lines,
respectively. In addition, there are 3 purely non-Abelian
diagrams, shown in Fig.3, whose coefficients will be denoted by
$t(n)$, $n=1,2,3$. The triangles are completely antisymmetric
with respect to the ordering of $F$'s, therefore they vanish for
the commuting generators.
\begin{figure}[h]\hfill
\includegraphics[scale=.2]{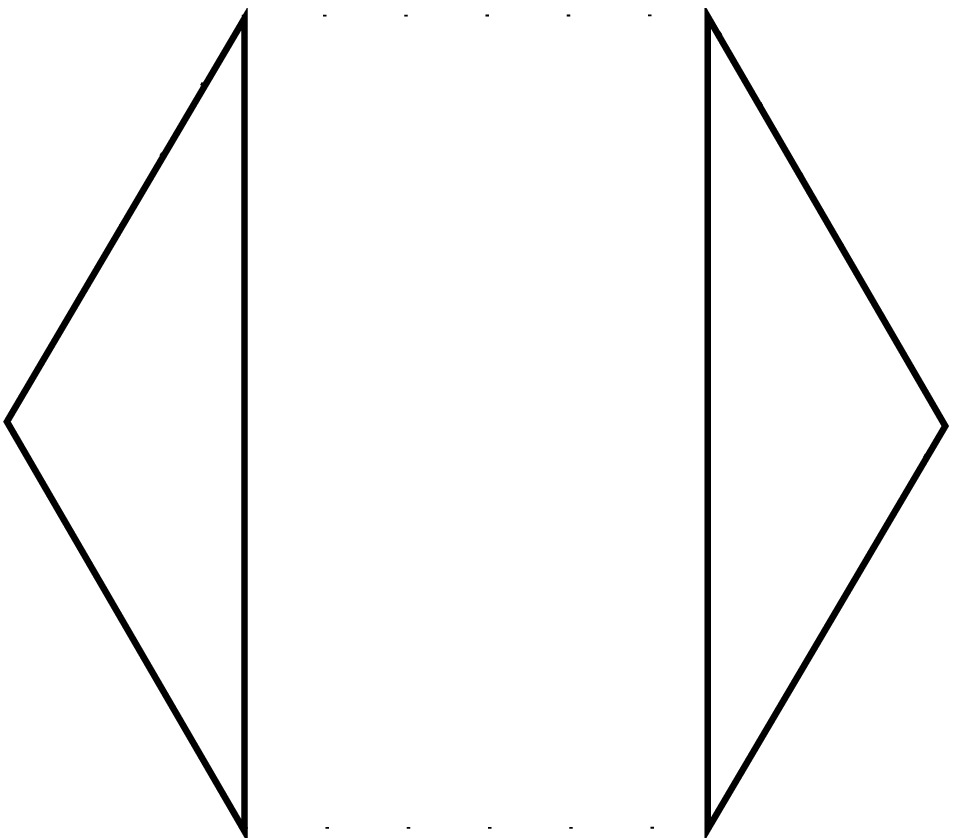}\hskip 1.5cm
\includegraphics[scale=.2]{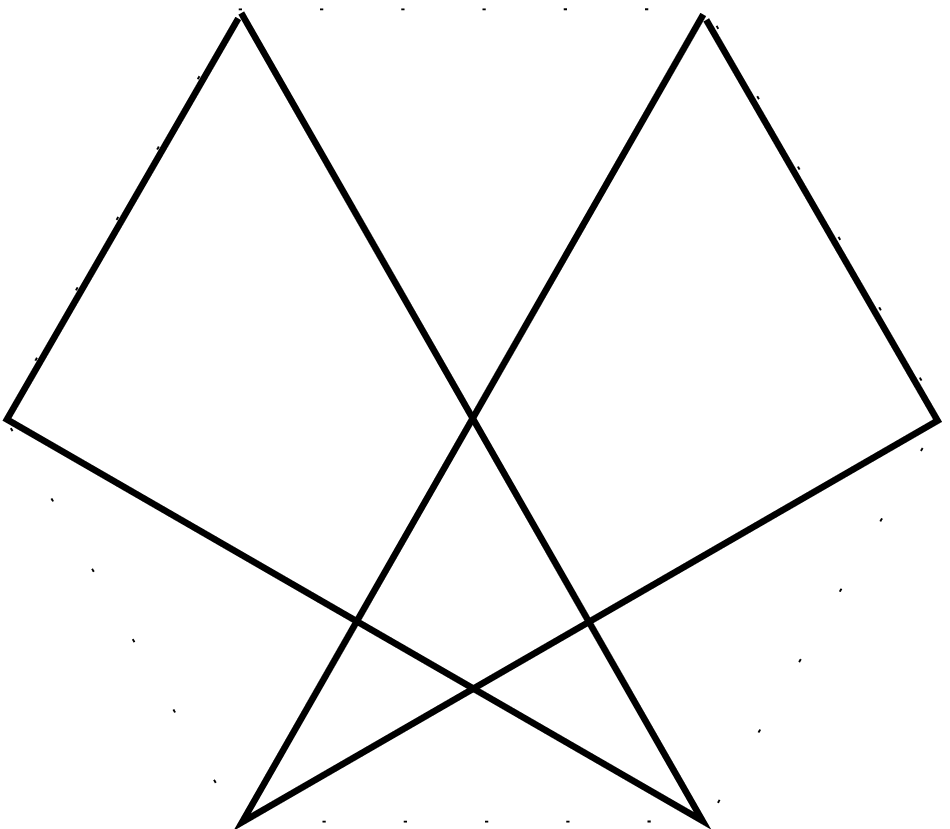}\hskip 1.5cm
\includegraphics[scale=.2]{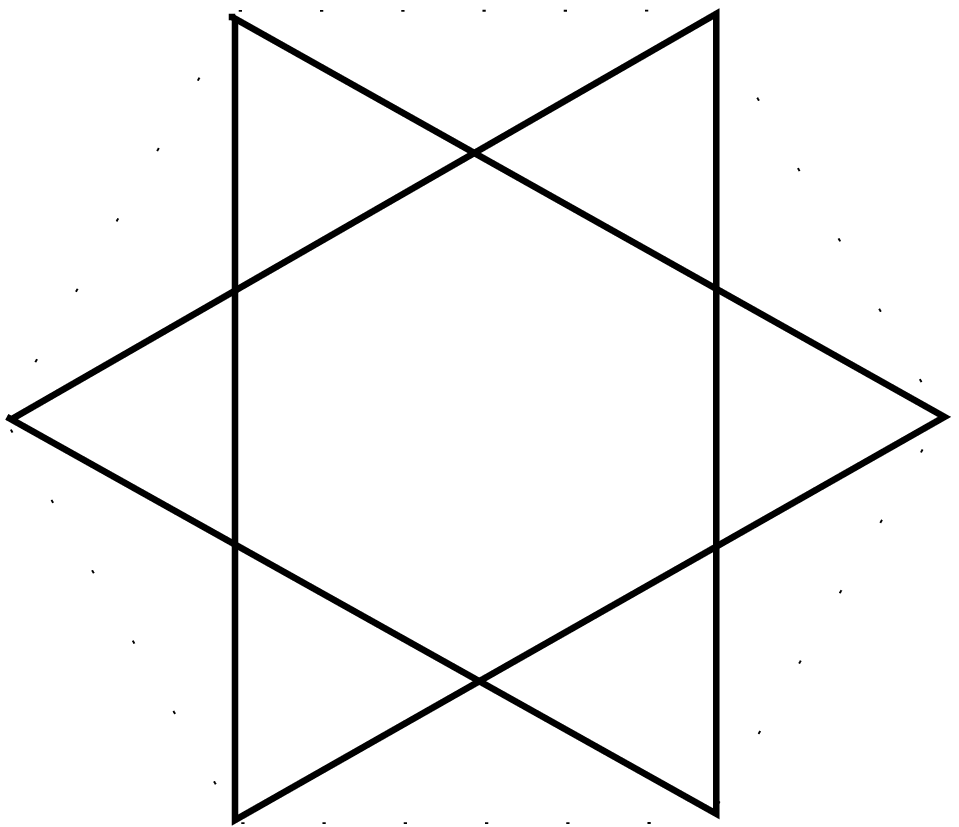}\hfill\hfill
\caption{Diagrams  representing three purely non-Abelian
terms. Their coefficients are, from left to right, $t(1)$, $t(2)$ and $t(3)$.}
\label{f3fig}
\end{figure}
\begin{center}
\begin{table}
\begin{tabular}{|c|c||c|c||c|c||c|c|}
\hline
Coeff & Perm&Coeff & Perm & Coeff & Perm & Coeff & Perm\\
\hline
$h(1)$ & $I$ & $h(8)$ & (132)(56)& $s(1)$ & $I$& $s(8)$ & (24)(36)  \\  \hline
$h(2)$ & (56) & $h(9)$ &  (12)(34)(56) & $s(2)$ & (45)& $s(9)$ & (24) \\  \hline
$h(3)$ & (13)(56) & $h(10)$ & (132)(45) & $s(3)$ & (34)& $l(1)$ & $I$  \\  \hline
$h(4)$ & (46) & $h(11)$ & (3564) &$ s(4)$ & (23)& $l(2)$ & (46)  \\  \hline
$h(5)$ & (13)(45) & $h(12)$ & (36) & $s(5)$ & (23)(45)& $l(3)$ & (24)  \\  \hline
$h(6)$ & (34)(56) & $h(13)$ & (364) & $s(6)$ & (23)(56)& $l(4)$ & (23)(56)  \\  \hline
$h(7)$ & (3465) & $h(14)$ & (23564) & $s(7)$ & (24)(56)& $l(5)$ & (34)  \\  \hline
\end{tabular}
\caption{List of permutations generating 28 Lagrangian terms
from three basic diagrams of Fig.2. ``Coeff" are
their coefficients while ``Perm'' are the generating
permutations in the standard cycle notation.}
\label{perms}
\end{table}
\end{center}

We will proceed from here in the same way as in the $F^4$ case.
We will identify a class of kinematic factors whose coefficients vanish
due to Riemann identity.
By comparing these zeroes with the corresponding field-theoretical amplitudes,
we will obtain a set of equations relating 31 coefficients of the effective
action.

In order to identify the vanishing coefficients we first consider the momenta
and polarization vectors of six gauge bosons distributed in 3
complex planes as follows:
\begin{equation}
\begin{tabular}{c}
$p_1~ p_2~ p_3~ p_4~ p_5~ p_6$\\ \hline\\[-2mm]
$\epsilon_1~ \epsilon_2$ \\ \hline\\[-2mm]
$\epsilon_3~ \epsilon_4~ \epsilon_5~ \epsilon_6$\\
 \hline
\end{tabular}\end{equation}
The vanishing coefficients are
\begin{equation}
[(p_1^+p_2^-p_3^+p_4^-p_5^+p_6^-)(\epsilon_1^-\epsilon_2^+)
(\epsilon_3^-\epsilon_4^+\epsilon_5^-
\epsilon_6^+)]~=~0\, \label{c1}
\end{equation}
and all its permutations in gauge boson indices (1,2,3,4,5,6).
Eq.(\ref{c1}) can be verified for both ``split'' and ``non-split'' contributions
by essentially the same methods as Eq.(\ref{final}), the only difference being
that now one has to deal with bigger chiral determinants. When all these
kinematic factors are compared with the amplitude computed by using
the most general
$F^6$ action, one finds that 17 combinations of the
Lagrangian coefficients must be zero. These combinations are listed in
Eq.(\ref{e4}) of Appendix A.

The second kinematic configuration is
\begin{equation}
\begin{tabular}{c}
$\epsilon_1~ \epsilon_5~ p_2~ p_6$ \\ \hline\\[-2mm]
$\epsilon_2~ \epsilon_4~ p_3~ p_5$\\ \hline\\[-2mm]
$\epsilon_3~ \epsilon_6~ p_1~ p_4$\\ \hline
\end{tabular}\end{equation}
Now the zero coefficients are
\begin{equation}
[(\epsilon_1^+\epsilon_5^+ p_2^- p_6^-)
(\epsilon_2^+\epsilon_4^+ p_3^- p_5^-)
(\epsilon_3^+\epsilon_6^+ p_1^- p_4^-)]~=~0\, \label{c2}
\end{equation}
and its permutations. After comparing with the general form
of the amplitude, we find 14 more vanishing combinations which are listed in
Eq.(\ref{e3}) of Appendix A.

Before discussing the solution of the combined system of constraints
consisting of Eqs.(\ref{e4}) and (\ref{e3}),
we would like to make a connection with the Abelian case. Note that
adding  Eqs.(\ref{e3}) yields
\begin{equation}
3\sum_{i=1}^{14}h(i)+8\sum_{j=1}^{9}s(i)  +32\sum_{k=1}^{5}l(k) =0\, ,
\label{ab2}
\end{equation}
while the sum of Eqs.(\ref{e4}) is
\begin{equation}
6\sum_{i=1}^{14}h(i)+24\sum_{j=1}^{9}s(i)  +96\sum_{k=1}^{5}l(k) =0\, .
\label{ab1}
\end{equation}
{}For the commuting generators, the non-Abelian action acquires the form of
(\ref{f6ab}), with
\begin{equation}
h=\sum_{i=1}^{14}h(i)\, ,\qquad s=\sum_{j=1}^{9}s(i)\, ,
\qquad l=\sum_{k=1}^{5}l(k)\, .\label{hsl1}
\end{equation}
Eqs.(\ref{ab2}-\ref{hsl1}) agree with Eqs.(\ref{33l}-\ref{hsl})
describing the $U(1)$
Cartan subalgebra sectors of the full $SO(32)$
theory. Thus we confirm our previous conclusion that the two-loop
effective action
of the heterotic
$[U(1)]^{16}$ gauge bosons is different from the Born-Infeld action; in particular,
the Lorentz hexagon term  disappears ($h=0$) in the Abelian limit.

It is a matter of simple
algebra to show that  Eqs.(\ref{e4}) and (\ref{e3})
contain only 19 independent combinations. Thus all effective action
coefficients can be parameterized in terms of 12 constants.
The  choice of the basis is completely arbitrary, however it is natural to
single out the purely non-Abelian coefficients $t(1)$, $t(2)$ and $t(3)$
and, for the remaining 9 constants, to select some combinations that are
directly related to certain scattering amplitudes. They are:
\begin{equation}
\begin{array}{c}
c(1)={4 l(1) + 6 l(2) + s(1)}\\
c(2)={4 l(1) + 2 l(5) + s(2)}\\
c(3)={6 l(2) + 2 l(5) + s(3)}\\
c(4)={4 l(5) + s(4)}\qquad\quad\;\\
c(5)={4 l(4) + 2 l(5) + s(5)}\\
c(6)={4 l(4) + 2 l(5) + s(6)}\\
c(7)={2 l(1) + 2 l(4) + s(7)}\\
c(8)={6 l(3) + 2 l(4) + s(8)}\\
c(9)={2 l(1) + 6 l(3) + s(9)}\end{array}\label{cis}\end{equation}
The problem of evaluating these constants will be discussed in Appendix B.
The corresponding amplitudes are complicated
because they involve nontrivial coupling between the left- and right-movers
and, unlike some tractable
higher genus amplitudes \cite{top},
they do not seem to have a simple topological origin.
In the present basis, the solution of Eqs.(\ref{e4}) and (\ref{e3}) reads
\begin{equation}
h=C_{h}\,c+T_{\!h}\,t\qquad s=C_{\! s}\,c+T_{\! s}\,t\qquad l=C_{l}\,
c+T_{l}\,t\label{sol1}
\end{equation}
where $C_{h,s,l}$ and $T_{h,s,l}$ are matrices whose explicit form is given in
Eqs.(\ref{ch}-\ref{tl}) of Appendix A.
As seen from the result, they are quite complicated, although it is
possible that a detailed analysis utilizing permutation theory
could reveal some interesting structure.
While we could not find a simple explanation for
the $c$-dependence,  the $t$-dependence
of the solution (\ref{sol1})
has an interesting interpretation that will be discussed below.

As mentioned before, the $F^6$ terms represented by the triangle diagrams
of Fig.3 disappear without a trace when one considers the Cartan subalgebra
sector, therefore they provide an interesting example of a genuine
non-Abelian structure. By using the antisymmetry of Lorentz
contractions, they can be rewritten  as
\begin{eqnarray}
{\cal L}_{6t}&=&t(1)\!
\stackrel{\circ}{F_{\beta_1\beta_2}}
\stackrel{\bullet}{F_{\alpha_1\alpha_2}}
\stackrel{\bullet}{F_{\alpha_2\alpha_3}}
\stackrel{\bullet}{F_{\alpha_3\alpha_1}}
\stackrel{\circ}{F_{\beta_2\beta_3}}
\stackrel{\circ}{F_{\beta_3\beta_1}}\nonumber\\
&&\hskip -3.6mm +~t(2)\!
\stackrel{\circ}{F_{\beta_1\beta_2}}
\stackrel{\bullet}{F_{\alpha_1\alpha_2}}
\stackrel{\bullet}{F_{\alpha_2\alpha_3}}
\stackrel{\circ}{F_{\beta_2\beta_3}}
\stackrel{\bullet}{F_{\alpha_3\alpha_1}}
\stackrel{\circ}{F_{\beta_3\beta_1}}\label{inf}\\
&&\hskip -3.2mm +~t(3)\!
\stackrel{\bullet}{F_{\alpha_1\alpha_2}}
\stackrel{\circ}{F_{\beta_1\beta_2}}
\stackrel{\bullet}{F_{\alpha_2\alpha_3}}
\stackrel{\circ}{F_{\beta_2\beta_3}}
\stackrel{\bullet}{F_{\alpha_3\alpha_1}}
\stackrel{\circ}{F_{\beta_3\beta_1}}\nonumber
\end{eqnarray}
where the circles over the gauge field strength matrices
mark antisymmetrization in their positions, in two separate sets
labeled by full and empty circles, respectively.
This can be represented by dressing the original diagrams with circles,
as shown in Fig.4.
\begin{figure}[h]\hfill
\includegraphics[scale=.2]{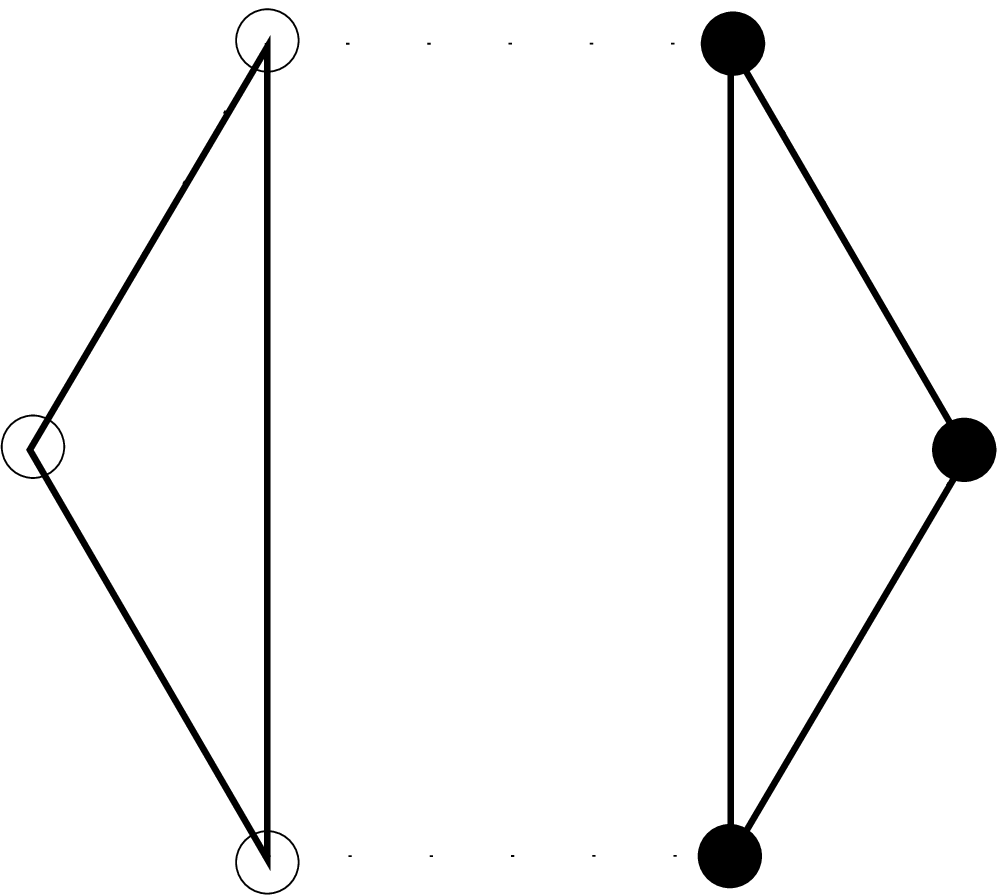}\hskip 1.5cm
\includegraphics[scale=.2]{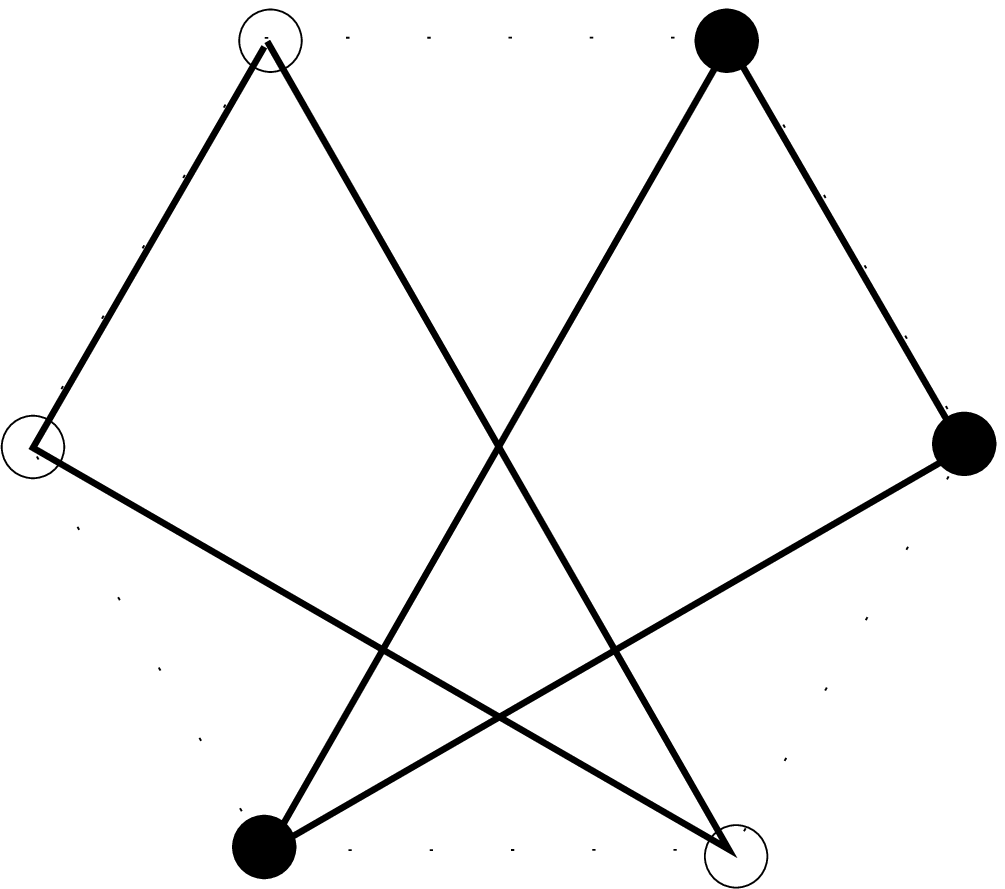}\hskip 1.5cm
\includegraphics[scale=.2]{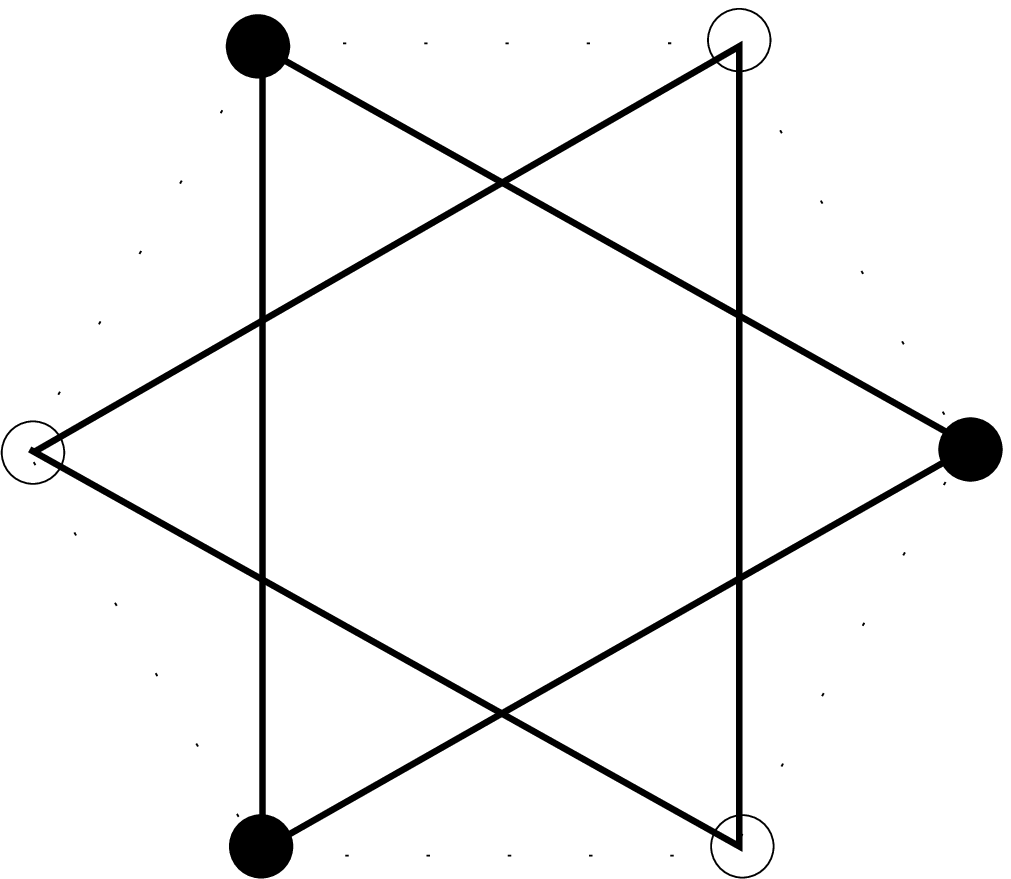}\hfill\hfill
\caption{The diagrams of Fig.3 are dressed  with circles which
label the two antisymmetrization sets.}
\label{f6t}
\end{figure}
According to Eqs.(\ref{sol1}), the relation of $h$, $s$, $l$
coefficients to $t(1)$, $t(2)$ and $t(3)$ is determined by the
matrices $T_{h,s,l}$ given in  Eqs.(\ref{th}-\ref{tl}). A closer
look at these matrices reveals a simple pattern. The Lagrangian
terms with three-line Lorentz contractions can be written as:
\begin{eqnarray}
{\cal L}_{6l}&=&-{3\over 4}t(1)\!
\stackrel{\circ}{F_{\alpha_1\alpha_2}}
\stackrel{\bullet}{F_{\alpha_2\alpha_1}}
\stackrel{\bullet}{F_{\beta_1\beta_2}}
\stackrel{\bullet}{F_{\gamma_1\gamma_2}}
\stackrel{\circ}{F_{\gamma_2\gamma_1}}
\stackrel{\circ}{F_{\beta_2\beta_1}}\nonumber\\
&&-{3\over 4}t(2)\!
\stackrel{\circ}{F_{\alpha_1\alpha_2}}
\stackrel{\bullet}{F_{\alpha_2\alpha_1}}
\stackrel{\bullet}{F_{\beta_1\beta_2}}
\stackrel{\circ}{F_{\gamma_2\gamma_1}}
\stackrel{\bullet}{F_{\gamma_1\gamma_2}}
\stackrel{\circ}{F_{\beta_2\beta_1}}\label{ltot}\\
&&-{3\over 4}t(3)\!
\stackrel{\bullet}{F_{\alpha_1\alpha_2}}
\stackrel{\circ}{F_{\alpha_2\alpha_1}}
\stackrel{\bullet}{F_{\beta_1\beta_2}}
\stackrel{\circ}{F_{\gamma_2\gamma_1}}
\stackrel{\bullet}{F_{\gamma_1\gamma_2}}
\stackrel{\circ}{F_{\beta_2\beta_1}}\, .\nonumber
\end{eqnarray}
They are  represented in a schematic way in Fig.\ref{f6l}, where the
antisymmetrizations are depicted
by inscribing the triangles of Fig.\ref{f6t} into the three-line diagram.
\begin{figure}[h]\hfill
\includegraphics[scale=.2]{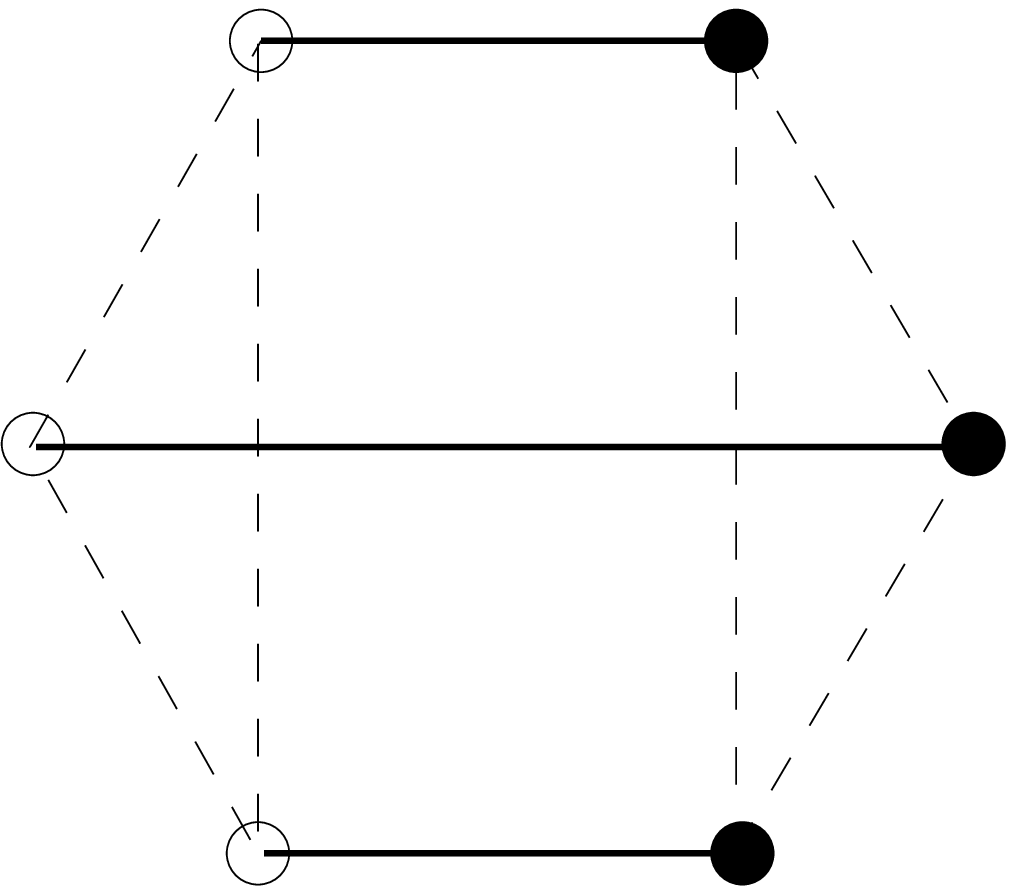}\hskip 1.5cm
\includegraphics[scale=.2]{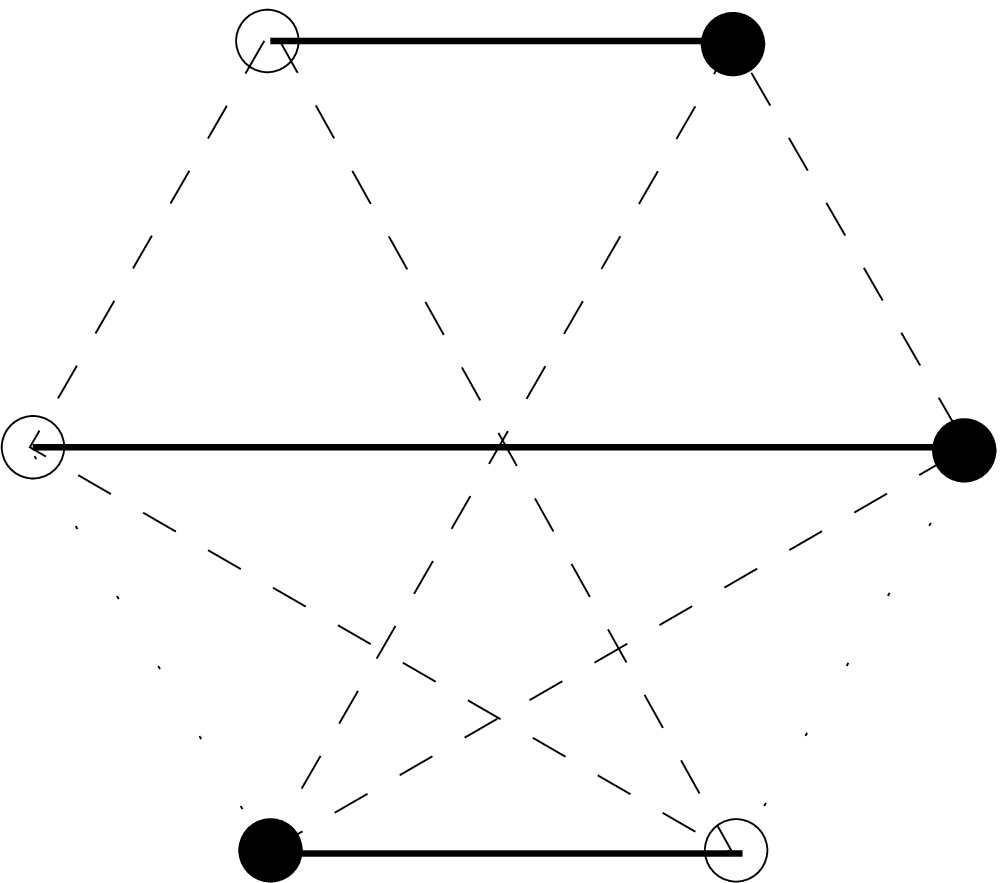}\hskip 1.5cm
\includegraphics[scale=.2]{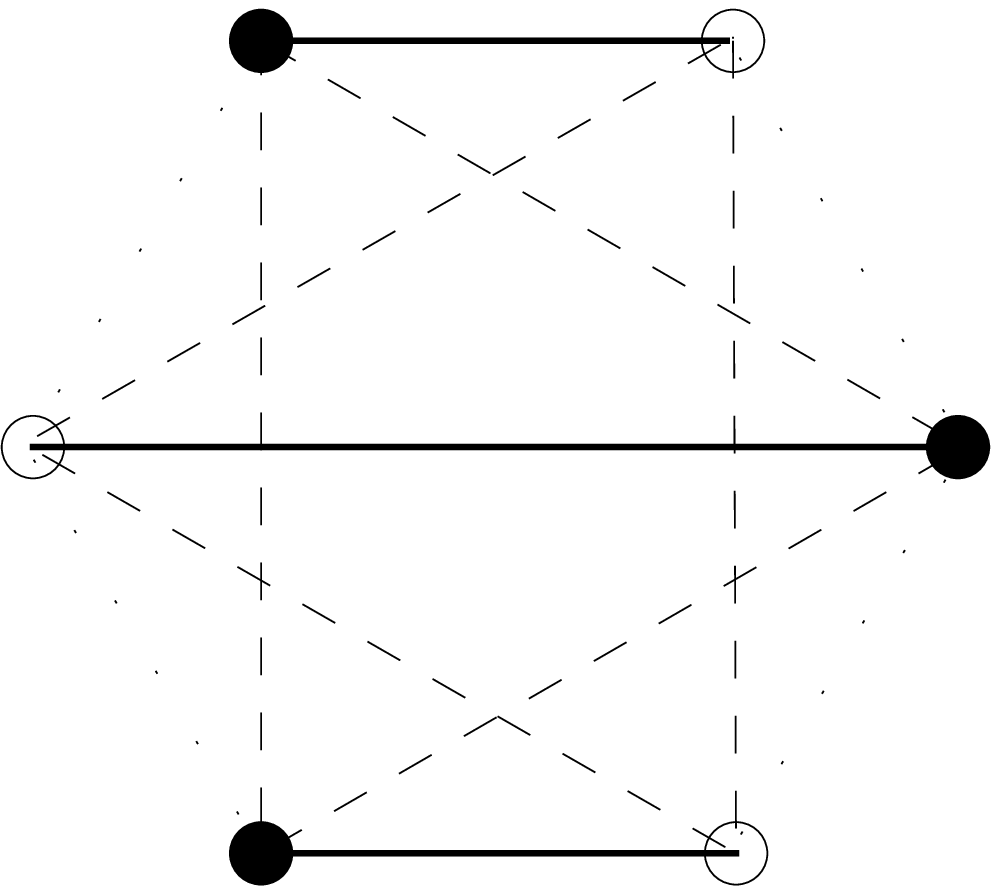}\hfill\hfill
\caption{Diagrammatic representation of ${\cal L}_{6l}$, Eq.(\ref{ltot}).}
\label{f6l}
\end{figure}
\begin{figure}[h]\hfill
\includegraphics[scale=.2]{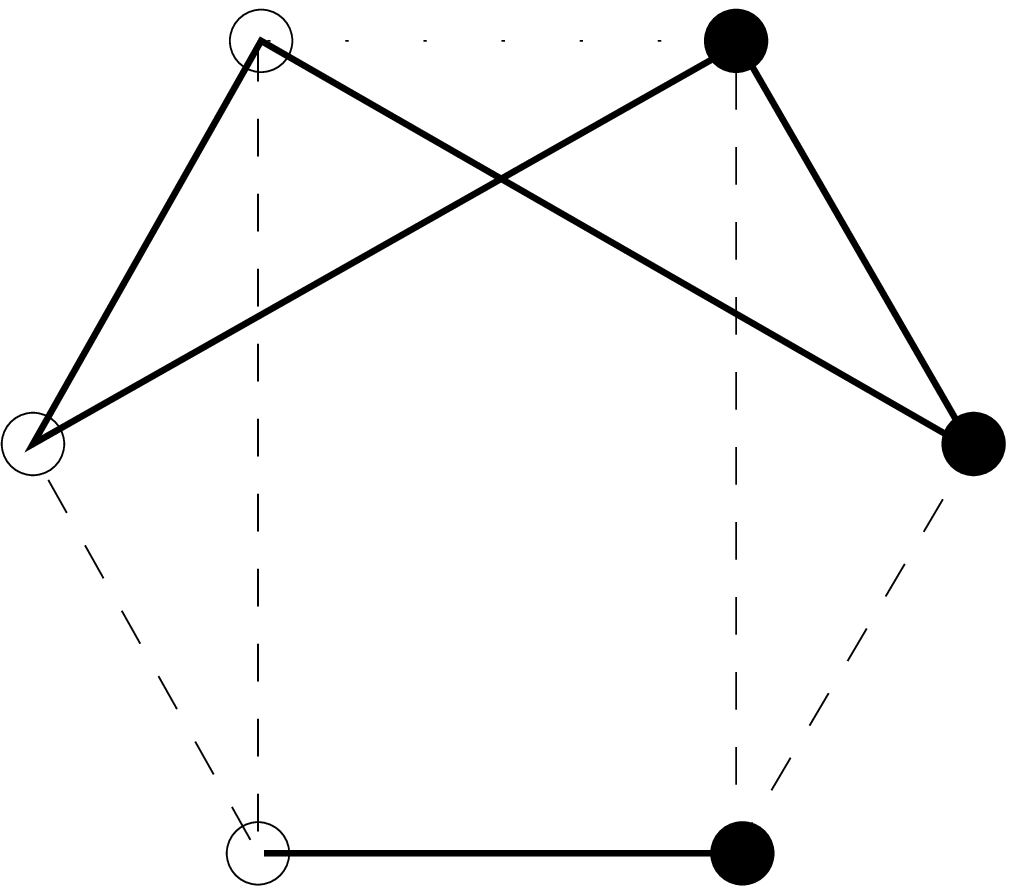}\hskip 1.5cm
\includegraphics[scale=.2]{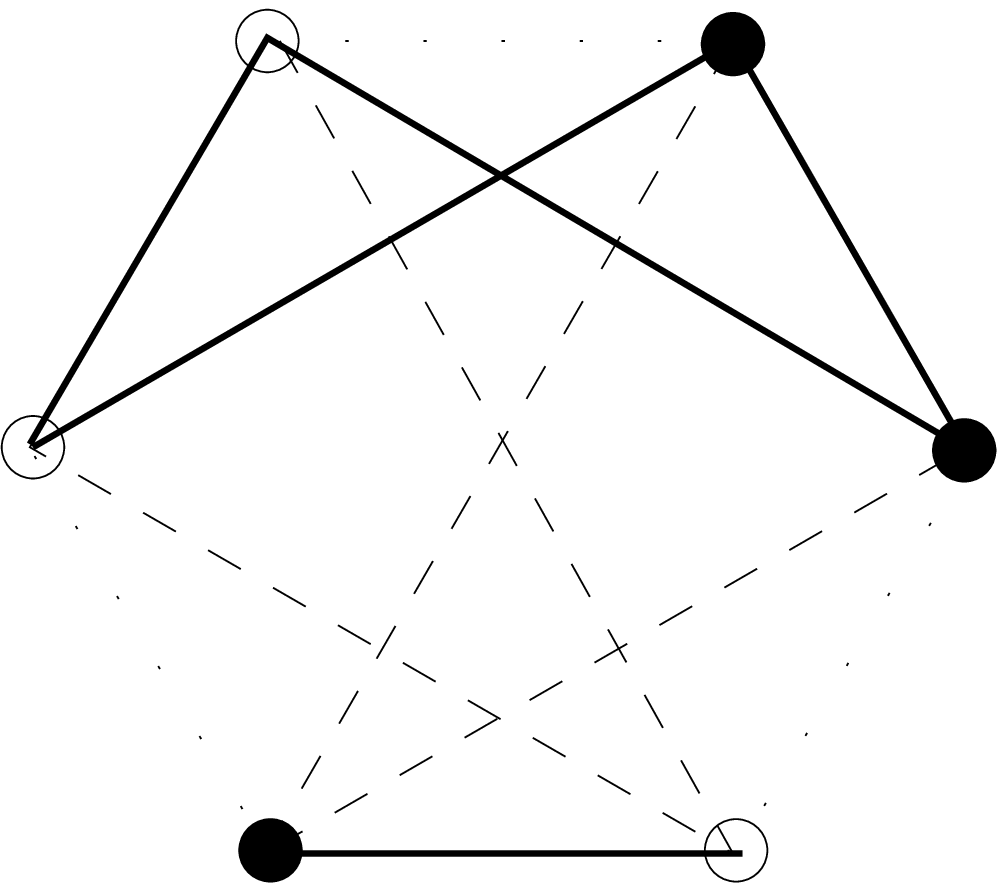}\hskip 1.5cm
\includegraphics[scale=.2]{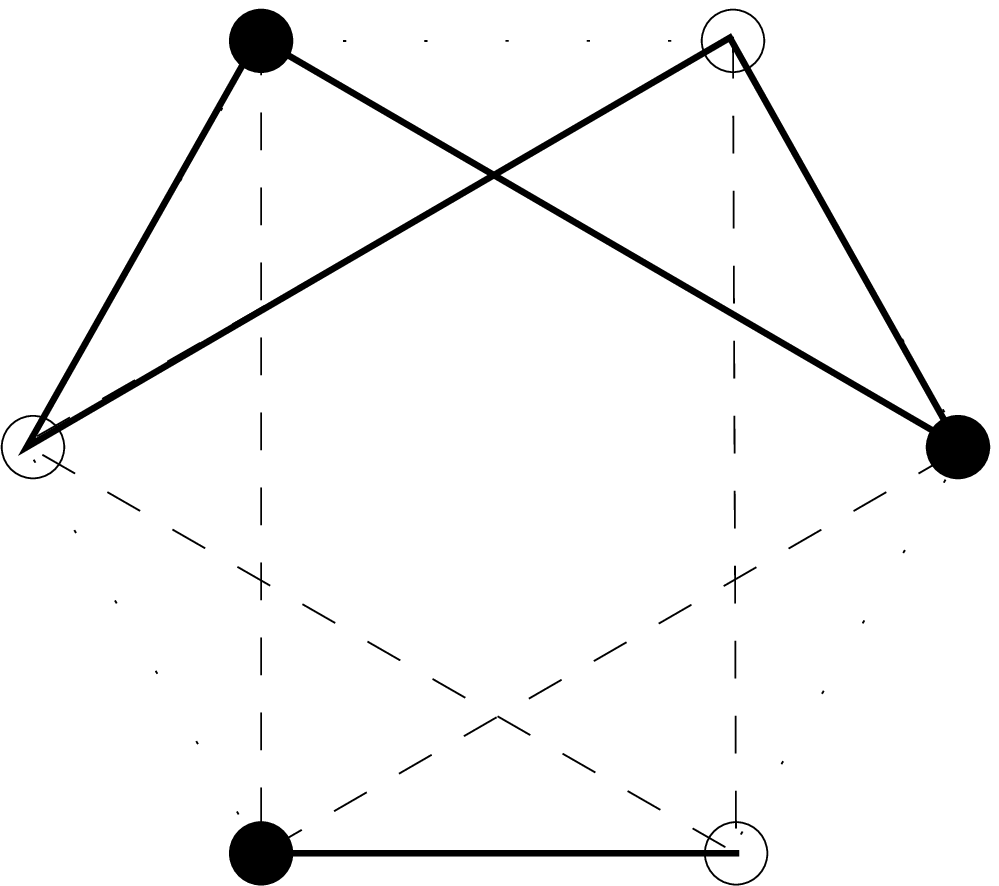}\hfill\hfill
\caption{Diagrammatic representation of ${\cal L}_{6s}$, Eq.(\ref{stot}).}
\label{f6s}
\end{figure}
The Lagrangian terms with Lorentz contractions related
to the square plus one line can be written as
\begin{eqnarray}
{\cal L}_{6s}&=&-{9\over 2}t(1)\!
\stackrel{\circ}{F_{\alpha_1\alpha_2}}
\stackrel{\bullet}{F_{\alpha_3\alpha_4}}
\stackrel{\bullet}{F_{\alpha_2\alpha_3}}
\stackrel{\bullet}{F_{\beta_1\beta_2}}
\stackrel{\circ}{F_{\beta_2\beta_1}}
\stackrel{\circ}{F_{\alpha_4\alpha_1}}\nonumber\\
&&-{9\over 2}t(2)\!
\stackrel{\circ}{F_{\alpha_1\alpha_2}}
\stackrel{\bullet}{F_{\alpha_3\alpha_4}}
\stackrel{\bullet}{F_{\alpha_2\alpha_3}}
\stackrel{\circ}{F_{\beta_1\beta_2}}
\stackrel{\bullet}{F_{\beta_2\beta_1}}
\stackrel{\circ}{F_{\alpha_4\alpha_1}}\label{stot}\\
&&-{9\over 2}t(3)\!
\stackrel{\bullet}{F_{\alpha_1\alpha_2}}
\stackrel{\circ}{F_{\alpha_3\alpha_4}}
\stackrel{\bullet}{F_{\alpha_2\alpha_3}}
\stackrel{\circ}{F_{\beta_1\beta_2}}
\stackrel{\bullet}{F_{\beta_2\beta_1}}
\stackrel{\circ}{F_{\alpha_4\alpha_1}}\, ,
\nonumber
\end{eqnarray}
see Fig.\ref{f6s}. Finally, there are two types of hexagon-like terms:
\begin{eqnarray}
{\cal L}_{6h}&=&-9t(1)\!
\stackrel{\circ}{F_{\alpha_1\alpha_2}}
\stackrel{\bullet}{F_{\alpha_2\alpha_3}}
\stackrel{\bullet}{F_{\alpha_3\alpha_4}}
\stackrel{\bullet}{F_{\alpha_4\alpha_5}}
\stackrel{\circ}{F_{\alpha_5\alpha_6}}
\stackrel{\circ}{F_{\alpha_6\alpha_1}}
\nonumber\\
&&-9t(2)\!
\stackrel{\circ}{F_{\alpha_1\alpha_2}}
\stackrel{\bullet}{F_{\alpha_2\alpha_3}}
\stackrel{\bullet}{F_{\alpha_3\alpha_4}}
\stackrel{\circ}{F_{\alpha_5\alpha_6}}
\stackrel{\bullet}{F_{\alpha_4\alpha_5}}\stackrel{\circ}{F_{\alpha_6\alpha_1}}
\label{htot1}\\
&&-9t(3)\!
\stackrel{\bullet}{F_{\alpha_1\alpha_2}}
\stackrel{\circ}{F_{\alpha_2\alpha_3}}
\stackrel{\bullet}{F_{\alpha_6\alpha_1}}
\stackrel{\circ}{F_{\alpha_4\alpha_5}}
\stackrel{\bullet}{F_{\alpha_5\alpha_6}}
\stackrel{\circ}{F_{\alpha_3\alpha_4}}\, ,
\nonumber
\end{eqnarray}
\begin{eqnarray}
{\cal L}_{6h'}&=&3t(1)\!
\stackrel{\circ}{F_{\alpha_1\alpha_2}}
\stackrel{\bullet}{F_{\alpha_2\alpha_3}}
\stackrel{\bullet}{F_{\alpha_6\alpha_1}}
\stackrel{\bullet}{F_{\alpha_4\alpha_5}}
\stackrel{\circ}{F_{\alpha_5\alpha_6}}
\stackrel{\circ}{F_{\alpha_3\alpha_4}}
\nonumber\\
&&\hskip -3mm +3t(2)\!
\stackrel{\circ}{F_{\alpha_1\alpha_2}}
\stackrel{\bullet}{F_{\alpha_2\alpha_3}}
\stackrel{\bullet}{F_{\alpha_6\alpha_1}}
\stackrel{\circ}{F_{\alpha_5\alpha_6}}
\stackrel{\bullet}{F_{\alpha_4\alpha_5}}\stackrel{\circ}{F_{\alpha_3\alpha_4}}
\label{htot2}\\
&&\hskip -3mm +3t(3)\!
\stackrel{\bullet}{F_{\alpha_1\alpha_2}}
\stackrel{\circ}{F_{\alpha_2\alpha_3}}
\stackrel{\bullet}{F_{\alpha_3\alpha_4}}
\stackrel{\circ}{F_{\alpha_4\alpha_5}}
\stackrel{\bullet}{F_{\alpha_5\alpha_6}}
\stackrel{\circ}{F_{\alpha_6\alpha_1}}\, .
\nonumber
\end{eqnarray}
They are represented in Figs.\ref{f6h} and \ref{f6hprime},
respectively.
Eqs.(\ref{inf}-\ref{htot2}) contain a closed
subset of $F^6$ Lagrangian terms in
$N{=}1$  supersymmetric $SO(32)$ gauge theory
describing the low-energy limit
of $D{=}10$ heterotic superstring.
\begin{figure}[h]\hfill
\includegraphics[scale=.2]{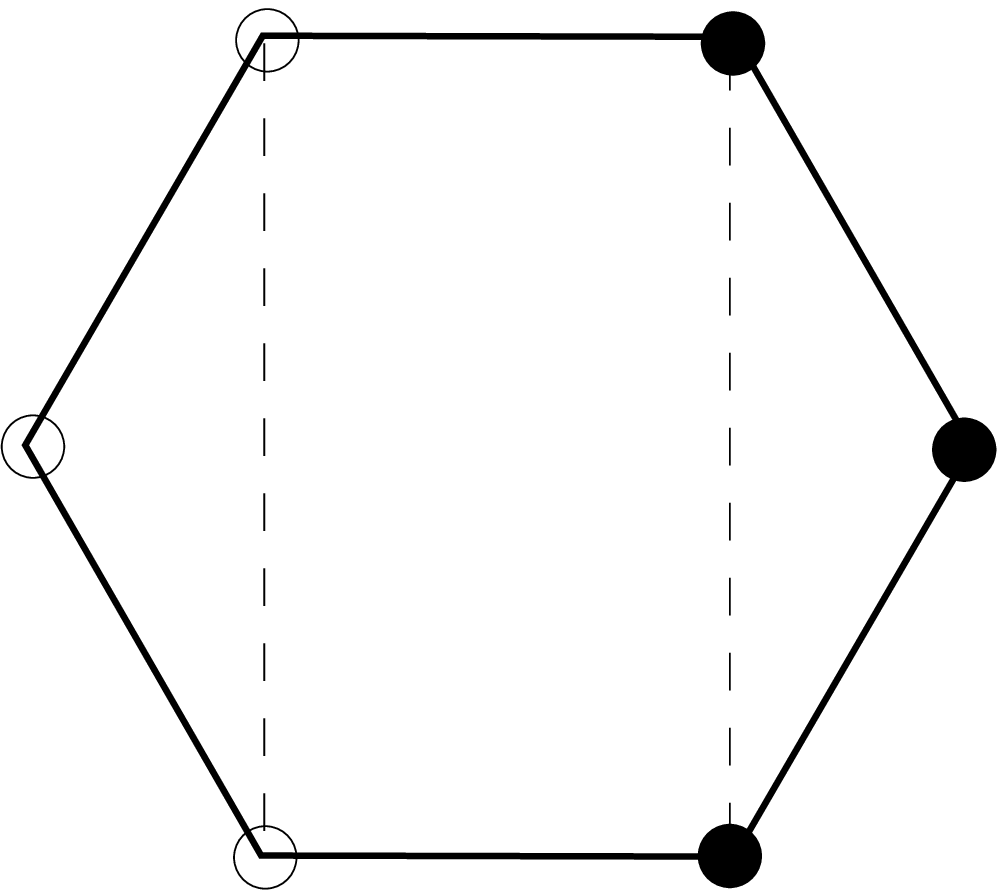}\hskip 1.5cm
\includegraphics[scale=.2]{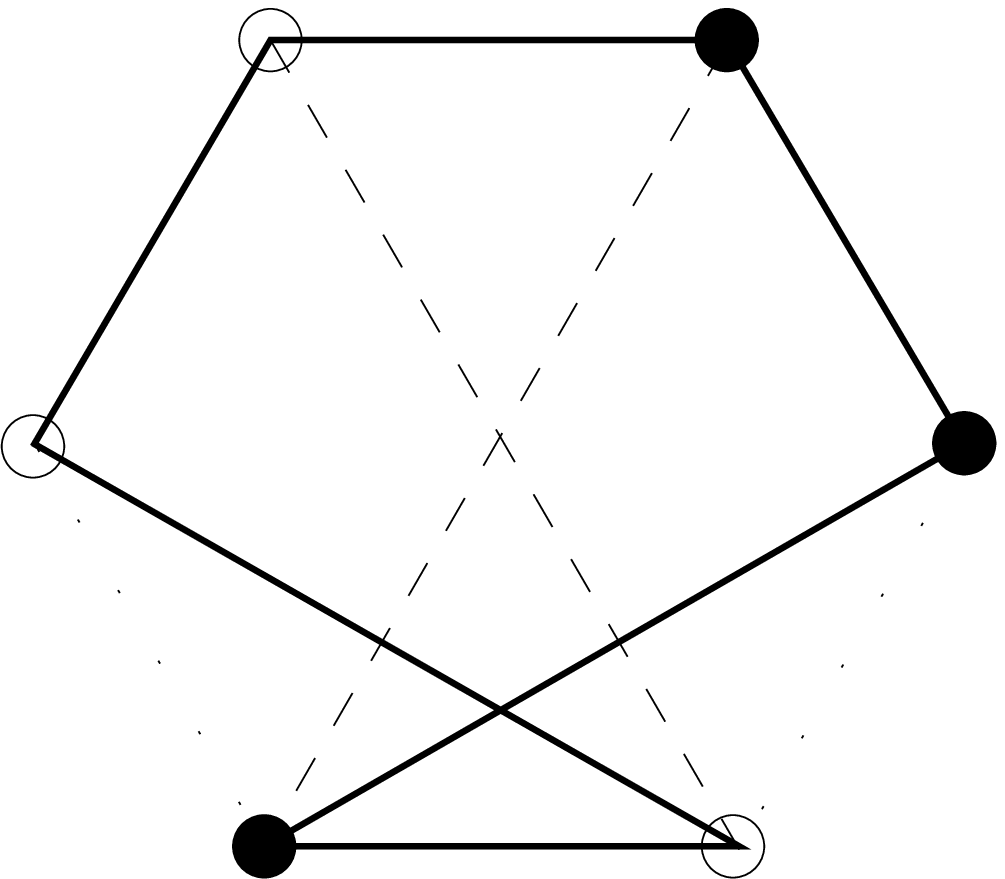}\hskip 1.5cm
\includegraphics[scale=.2]{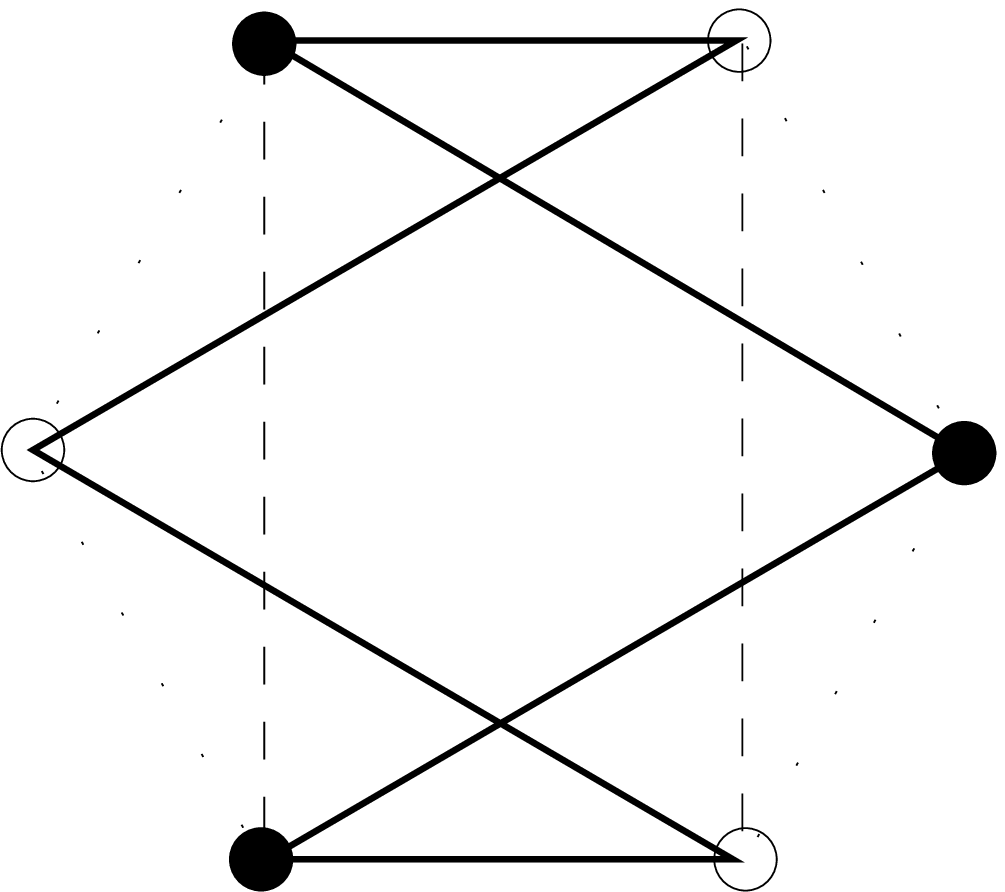}\hfill\hfill
\caption{Diagrammatic representation of ${\cal L}_{6h}$, Eq.(\ref{htot1}).}
\label{f6h}
\end{figure}
\begin{figure}[h]\hfill
\includegraphics[scale=.2]{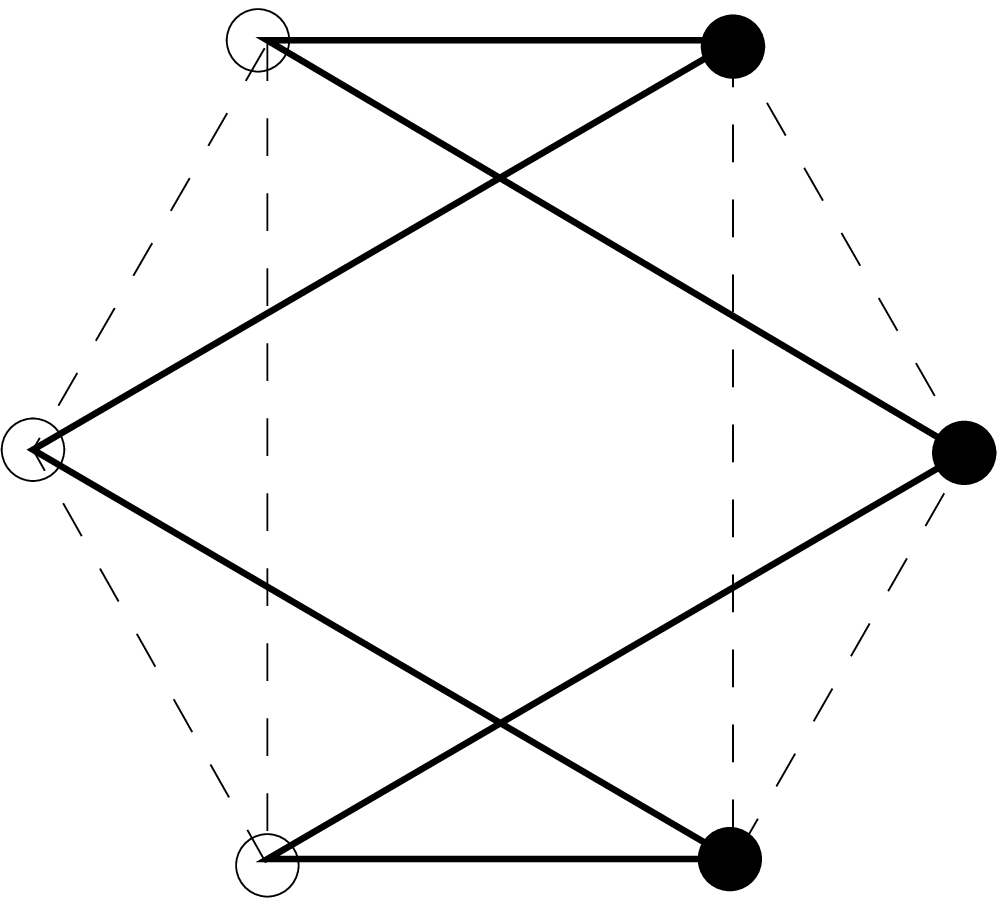}\hskip 1.5cm
\includegraphics[scale=.2]{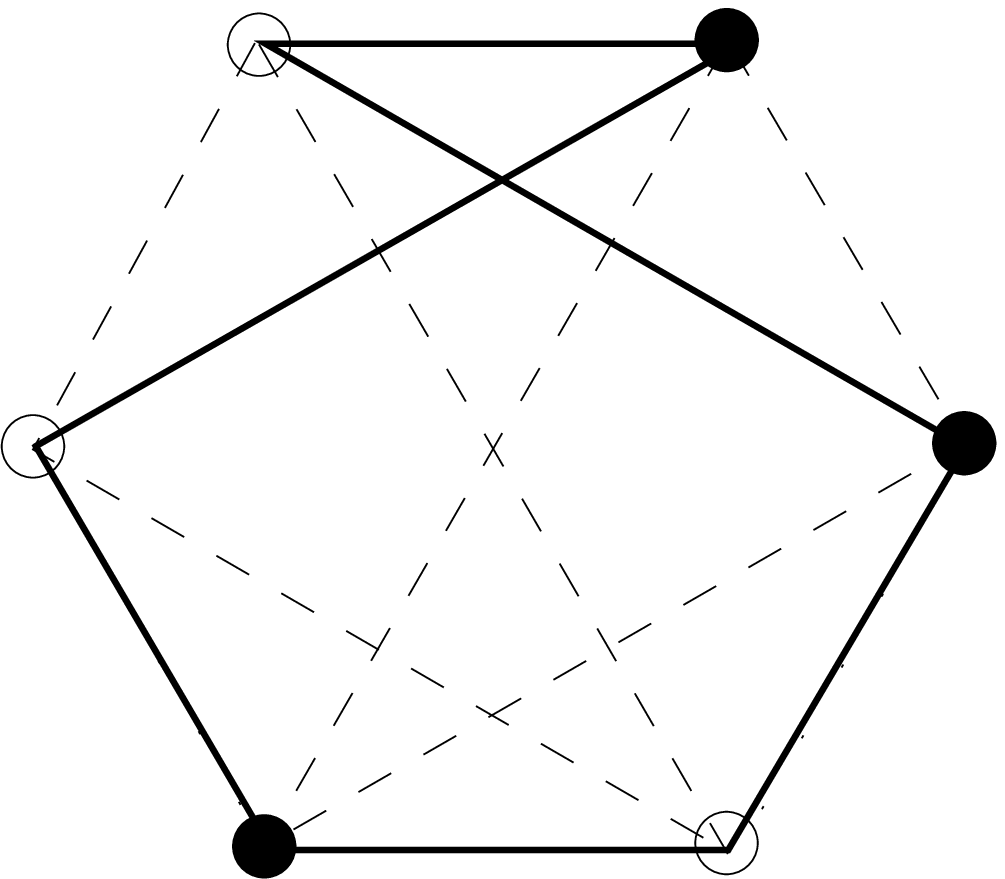}\hskip 1.5cm
\includegraphics[scale=.2]{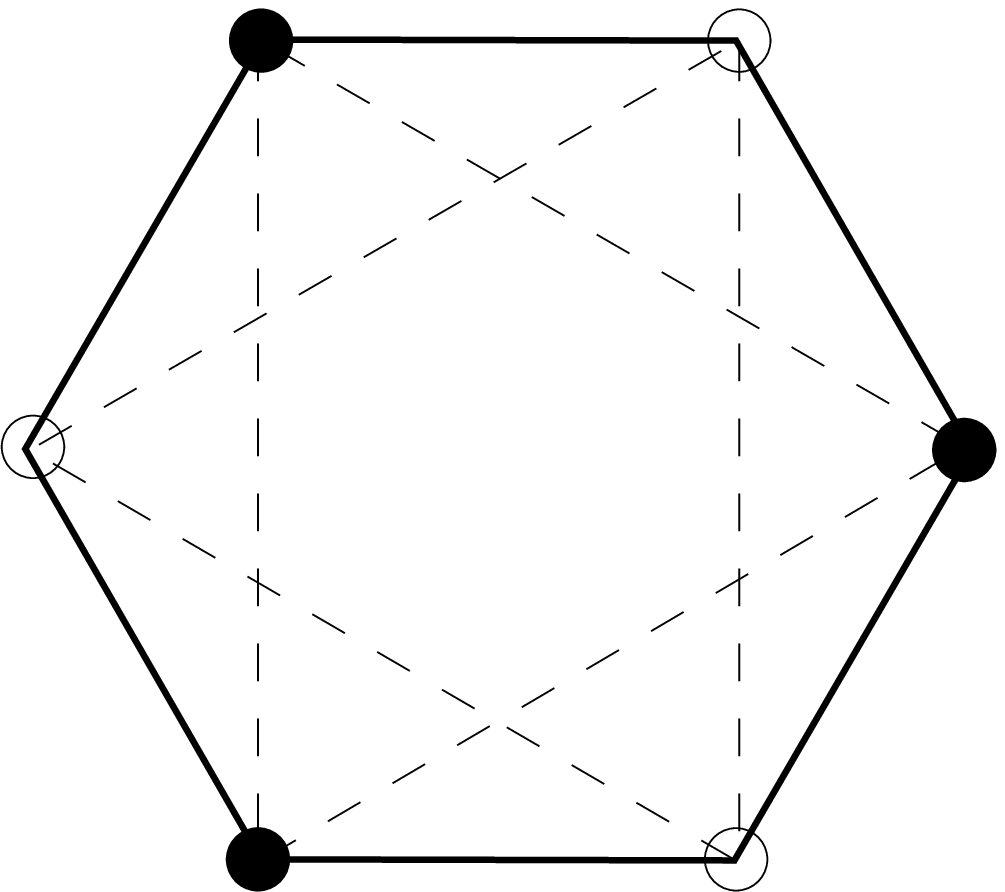}\hfill\hfill
\caption{Diagrammatic representation of ${\cal L}_{6h'}$, Eq.(\ref{htot2}).}
\label{f6hprime}
\end{figure}

A technical comment is in order here.
We derived our equations in the unitary
gauge \req{choice}.
{}For a different choice of the PCO insertion points $(x_1,x_2)$,
the correlation functions could change by a
total derivative term
(w.r.t.\ to the moduli space of the genus two Riemann surface) \cite{Verlinde:1987sd}.
The latter would
vanish after integration over the moduli space provided there is no contribution
from the boundary.
We checked that our equations do not receive corrections from such boundary
terms.
Their corresponding factorization limits vanish as a consequence of Riemann identities.
Thus they are valid for any gauge choice.

\section{Conclusions}
When a new theory is formulated, it is important not only to verify
its content, but also to test its limitations.
Our original motivation for the two-loop heterotic computations was to
obtain the Non-Abelian Born-Infeld action for Type I superstring theory.
Na\"{\i}vely, this looked like a straightforward
application of superstring duality \cite{Polchinski:1995df,apt}
but the results turned out to be quite surprising. Once we managed to
determine the two-loop heterotic $F^6$ terms, we found that their structure
is fundamentally different from any form of Born-Infeld generalization.
Namely, when the gauge group generators were restricted to the Cartan subalgebra,
we obtained an  action which did not agree with the weak
field expansion of Abelian Born-Infeld theory. Before offering some explanations,
we would like to recall the origin of this result.

Our conclusions rely on the Riemann theta function
identity (\ref{riemann}) which
reflects supersymmetry in superstring scattering amplitudes. The disagreement with
the Abelian Born-Infeld action first appears  in Eqs.(\ref{33l}) and (\ref{46l})
as a consequence of this identity. To be more precise, while Eq.(\ref{46l})
is still consistent with the Born-Infeld values $H{=}L{=}{-}S{=}1$, \linebreak
Eq.(\ref{33l}) is {\em
not}. The latter reappears later at the non-Abelian level as Eq.(\ref{ab2}), the
sum of 14 Eqs.(\ref{e3}) describing the vanishing kinematic
factors (\ref{c2}). Hence, as far as superstring amplitudes are
concerned,
Eq.(\ref{c2}) is technically
responsible for a conflict with Born-Infeld theory.

Looking at our efforts in a broader perspective,
it becomes clear that the $F^6$
interaction terms have a different character than the well-known $F^4$ terms.
As shown in Section 2, the   Tr$(t_8F^4)$ structure
follows from supersymmetry, and the scattering amplitude that determines
the overall normalization is a ``BPS-saturated''
amplitude which has a topological character. It is related
to Green-Schwarz anomaly; the two-loop contribution vanishes
in agreement with the well-established non-renormalization theorems
\cite{part1}. The two-loop heterotic
$F^6$ terms are also constrained by supersymmetry.
While the most general ansatz contains 31 constants, the use of Riemann
identity allows parameterizing it by only 12 constants.
In Section 4, we discussed some properties of this effective action.
What makes $F^6$ truly different from
$F^4$ is that the corresponding amplitudes are {\em not\/}
``BPS-saturated'' -- they exhibit non-trivial couplings
between the left- and right-movers, as shown explicitly in Appendix
B. Thus the $F^6$ terms are sensitive to the full spectrum of
superstring theory and the comparison of dual descriptions
may be quite subtle. So far all investigations of string amplitudes in the context of
duality symmetries have been limited to
couplings where only short (and intermediate \cite{LS}) multiplets contribute.
Understanding how superstring duality
works for generic amplitudes would certainly improve our understanding
of superstring dynamics.\\[1cm]
{\bf Acknowledgements}\vskip 5mm

We are grateful to Alexander Sevrin and Jan Troost for stimulating discussions
and correspondence.
\appendix
\section{Solution}

The 17 linear combinations of Lagrangian coefficients that vanish as a
consequence of Eq.(\ref{c1}) are:
\begin{equation}\hskip -.7cm
\begin{array}{c}
{6 h(1) + 4 h(4) + 3 h(9)  + 16 l(1) + 24 l(2)+ 10 s(1) + 4 s(9)}\\
{h(2) + h(3) + h(6) + h(8) + 24 l(2) + 8 l(5) + 2 s(1) + 6 s(3) + 2 s(4) + 2 s(6)}\\
{h(2) + h(3) + h(5) + h(7) + 16 l(1) + 8 l(5) + 2 s(1) + 4 s(2) + 4 s(4) + 4 s(7)}\\
{h(2) + h(5) + h(6) + h(8) + 24 l(2) + 8 l(5) + 2 s(1) + 6 s(3) + 2 s(4) + 2 s(5)}\\
{h(2) + h(3) + h(5) + h(6) + 24 l(2) + 16 l(1) + 4 s(1) + 2 s(2) + 4 s(3) + 4 s(7)}\\
{2 h(10) + h(11) + 2 h(12) + 2 h(13)
+ 16 l(1) + 8 l(5) + 6 s(2) + 2 s(5) + 2 s(6) + 4 s(9)}\\
{h(2) + h(3) + h(7) + h(8) + 16 l(5) + 2 s(2) + 2 s(3) + 6 s(4) + 2 s(6)}\\
{h(11) + 2 h(12) + 2 h(13) + 2 h(14)
+ 16 l(4) + 8 l(5) + 2 s(2) + 2 s(5) + 6 s(6) + 4 s(8)}\\
{h(5) + h(6) + h(7) + h(8) + 16 l(4)
+ 8 l(5) + 2 s(3) + 2 s(4) + 4 s(5) + 2 s(6) + 4 s(7)}\\
{h(3) + h(6) + h(7) + h(8) + 16 l(4)
+ 8 l(5) + 2 s(3) + 2 s(4) + 2 s(5) + 4 s(6) + 4 s(7)}\\
{h(2) + h(5) + h(7) + h(8) + 16 l(5) + 2 s(2) + 2 s(3) + 6 s(4) + 2 s(5)}\\
{2 h(10) + h(11) + 2 h(12) + 2 h(14)
+ 16 l(4) + 8 l(5) + 2 s(2) + 6 s(5) + 2 s(6) + 4 s(8)}\\
{h(3) + h(5) + h(6) + h(7) + 16 l(1)
+ 16 l(4) + 2 s(1) + 2 s(2) + 2 s(5) + 2 s(6) + 8 s(7)}\\
{h(10) + h(11) + h(13) + 8 l(1) + 24 l(3) + 2 s(2) + 4 s(8) + 4 s(9)}\\
{h(10) + h(11) + h(14) + 24 l(3) + 8 l(4) + 2 s(5) + 6 s(8) + 2 s(9)}\\
{2 h(4) + 3 h(9) + 8 l(1) + 24 l(3) + 2 s(1) + 8 s(9)}\\
{h(11) + h(13) + h(14) + 24 l(3) + 8 l(4) + 2 s(6) + 6 s(8) + 2 s(9)}
\end{array}\label{e4}\end{equation}

The 14 combinations that vanish as a consequence of Eq.(\ref{c2}) are
\begin{equation}
\begin{array}{c}
{3 h(1) + h(6) + 8 l(1) + 2 s(1) + 2 s(7) + t(1)}\\
{2 h(2) + h(8) + 8 l(5) + 2 s(3) + 4 s(4) - 2 t(1)}\\
{h(3) + h(12) + 8 l(4) + 2 s(6) + 2 s(7) + t(1)}\\
{h(5) + h(12) + 8 l(4) + 2 s(5) + 2 s(7) + t(1)}\\
{h(4) + h(10) + h(13) + 24 l(3) + 2 s(8) + 4 s(9) - t(1)}\\
{h(2) + 3 h(9) + h(11) + 16 l(1) + 2 s(1) + 2 s(2) + 4 s(9) + t(2)}\\
{2 h(6) + 2 h(14) + 8 l(5) + 2 s(3) + 2 s(5) + 2 s(6) - t(2)}\\
{h(3) + 2 h(4) + h(5) + 24 l(2) + 4 s(1) + 2 s(3) - t(2)}\\
{h(5) + h(7) + 2 h(13) + 8 l(5) + 2 s(2) + 2 s(4) + 2 s(6) + t(2)}\\
{h(3) + h(7) + 2 h(10) + 8 l(5) + 2 s(2) + 2 s(4) + 2 s(5) + t(2)}\\
{h(8) + 2 h(11) + 16 l(4) + 2 s(5) + 2 s(6) + 4 s(8) - t(2)}\\
{h(7) + h(12) + 8 l(1) + 2 s(2) + 2 s(7) + 3 t(3)}\\
{h(14) + 8 l(3) + 2 s(8) - t(3)}\\
{h(8) + 8 l(2) + 2 s(3) - 2 t(3)}
\end{array}\label{e3}\end{equation}

The matrices describing the solution (\ref{sol1})
of the full set of constraints are given by:
\begin{equation}
C_h=\left(\!\!
\begin{array}{ccccccccc}
    0&0&1&-1&0&0&0&0&0 \\
    0&0&2&0&0&0&-4&0&0 \\
    0&2&-1&0&1&-1&0&-2&0 \\
    0&2&0&-1&-1&-1&2&0&0 \\
    0&2&-1&0&-1&1&0&-2&0 \\
    2&-2&1&-2&1&1&0&2&-4 \\
    0&0&-1&0&1&1&0&-2&0 \\
    0&0&0&2&0&0&-4&0&0 \\
    0&0&1&0&-1&-1&0&2&0 \\
    0&0&-1&0&-1&1&2&0&0 \\
    -2&-2&1&2&-1&-1&0&2&4 \\
    0&0&-1&1&0&0&0&0&0 \\
    0&0&-1&0&1&-1&2&0&0 \\
    0&-2&0&-1&1&1&2&0&0
\end{array}\!\right)\label{ch}
\end{equation}
\begin{equation}
C_s=\frac{1}{2}\left(\!\!
\begin{array}{ccccccccc}
    -2&-2&-3&4&1&1&-2&0&2 \\
    0&-4&3&-2&1&1&-2&4&-2 \\
    -2&2&-4&0&0&0&6&-2&2 \\
    2&-2&0&-4&0&0&6&2&-2 \\
    0&2&1&0&-1&-3&-2&0&2 \\
    0&2&1&0&-3&-1&-2&0&2 \\
    -1&-1&2&2&-1&-1&-3&1&1 \\
    1&3&-1&0&0&0&-1&-3&-3 \\
    1&-1&0&-1&2&2&-1&-3&-3
\end{array}\!\right)\label{cs}
\end{equation}
\begin{equation}
C_l=\frac{1}{8}\left(\!\!
\begin{array}{ccccccccc}
    2&-2&4&-2&0&0&-2&2&-2 \\
    1&5&-3&-1&-1&-1&5&-3&1 \\
    -1&-1&1&1&-1&-1&-1&3&3 \\
    1&-3&-1&-3&3&3&5&1&-3 \\
    -2&2&0&6&0&0&-6&-2&2
\end{array}\!\right)\label{cl}
\end{equation}\vskip 2mm
\begin{equation}
T_h=\frac{1}{2}\left(\!\!\!
\begin{array}{cccccccccccccc}
    -1&4&-2&1&-2&-2&0&2&0&0&2&-1&0&-1 \\
    0&0&-1&1&-1&3&-1&0&-1&-1&1&0&-1&1 \\
    1&0&0&-3&0&0&-6&6&2&0&0&-3&0&3
\end{array}\!\right)^{\!\!\sim}\label{th}
\end{equation}

\begin{equation}
T_s=\frac{1}{4}\left(\!\!\!
\begin{array}{ccccccccc}
    2&0&-2&-4&2&2&3&-1&-2 \\
    -1&3&-2&2&-1&-1&0&-1&1 \\
    0&6&-6&0&0&0&3&-3&0
\end{array}\right)^{\!\!\sim}\label{ts}
\end{equation}\vskip 2mm

\begin{equation}
T_l=\frac{1}{8}\left(
\begin{array}{ccccc}
    0&-1&1&-2&2 \\
    1&-1&0&1&-1 \\
    2&-3&1&0&0
\end{array}\!\right)^{\!\!\!\sim}\label{tl},
\end{equation}
where the tilde symbol denotes matrix transposition.
\section{Evaluation of the  Amplitudes}
The coefficients $c(n)$, $n=1,\dots,9$, defined in
Eqs.(\ref{cis}) which, together with $t(k)$,
fully determine the two-loop heterotic $F^6$ action,
are directly related to certain scattering amplitudes.
In order to illustrate some general features
of these amplitudes, we discuss here
one specific class of contributions to $c(1)$. This coefficient can be
extracted by considering the following kinematic configuration:
\begin{equation}
\begin{tabular}{c}
$\epsilon_1~ \epsilon_2~ p_1~ p_2$ \\ \hline\\[-2mm]
$\epsilon_3~ \epsilon_4~ \epsilon_5~ \epsilon_6 $\\ \hline\\[-2mm]
$p_3~ p_4~ p_5~ p_6$\\ \hline
\end{tabular}\label{conf2}
\end{equation}
Then
\begin{equation} c(1)~=~4[(p_1\!\cdot\! p_2)(\epsilon_1\! \cdot\!\epsilon_2)
(p_3^+p_4^-p_5^+p_6^-)(\epsilon_3^-\epsilon_4^+
\epsilon_5^-\epsilon_6^+)] .\label{final1}
\end{equation}
There are many ``split'' and
``non-split'' contributions to this coefficient. For the moment, we
focus on the ``non-split''  correlator
\begin{eqnarray}
A(z,x)\!&=&\!\langle \bar{\psi}(z_1)\psi(z_2)\rangle^2
\langle
\bar{\psi}(z_3)\bar{\psi}
(z_5)\psi(z_4)\psi(z_6)\rangle\nonumber\\
&&\,\times\langle
\bar{\psi}(z_4)\bar{\psi}(z_6)\bar{\psi}(x_1)\psi(z_3)
\psi(z_5)\psi(x_2)\rangle\, ,\label{azx}
\end{eqnarray}
involving the PCO fermions associated to the $(p_3,p_4,p_5,p_6)$ plane.
It yields the sum
\begin{equation}
\dots\sum_{\delta\,\makebox{\scriptsize even}}
\langle\alpha|\delta\rangle\,
\theta_{\delta}^2(z_1{-}z_2)
\theta_{\delta}(z_4{+}z_6{+}x_1{-}z_3{-}z_5{-}x_2)
\theta_{\delta}(z_3{+}z_5{-}z_4{-}z_6) ,
\end{equation}
which, upon applying the Riemann identity (\ref{riemann}), gives
\begin{eqnarray}&&{\dots E^2(z_3,z_5)\over\dots E^2(z_1,z_2)}\,
\theta_{\alpha}(z_1{+}x_1{-}z_2{-}\Delta_{\alpha})
\theta_{\alpha}(z_2{+}x_1{-}z_1{-}\Delta_{\alpha})\nonumber\\ &&\times\,
\theta_{\alpha}(z_4{+}z_6{+}x_1{-}z_3{-}z_5{-}\Delta_{\alpha})
\theta_{\alpha}(z_3{+}z_5{+}x_2{-}z_4{-}z_6{-}\Delta_{\alpha})\, .
\end{eqnarray}
Theta functions of such arguments are related to
the  correlators of spin-1 $b$-$c$ system \cite{Verlinde:1986kw}.
By using the bosonisation
formula, we obtain
\begin{eqnarray}
A(z,x)\!&=&\dots
\left|\begin{array}{cc}\omega_1(z_1)\!&\! \omega_2(z_1)\\
\omega_1(x_1)\!&\! \omega_2(x_1)\end{array}\right|
\left|\begin{array}{cc}\omega_1(z_2)\!&\! \omega_2(z_2)\\
\omega_1(x_1)\!&\! \omega_2(x_1)\end{array}\right|\nonumber\\
&&\hskip -2cm\times
\left|\begin{array}{cc}\omega_1(z_3)\!&\! \omega_2(z_3)\\
\omega_1(z_5)\!&\! \omega_2(z_5)\end{array}\right|
\left|\begin{array}{cc}\omega_1(z_4)\!&\! \omega_2(z_4)\\
\omega_1(z_6)\!&\! \omega_2(z_6)\end{array}\right|
\Lambda_{z_3}(x_1,z_5)\Lambda_{z_4}(x_2,z_6)\, ,\label{azxfin}
\end{eqnarray}
where $\omega_{1,2}$ inside determinants
are the canonical basis of holomorphic 1-forms
and $\Lambda_p(x,y)$ are Cauchy kernels \cite{Fay}.
The prefactor represented by
the ellipsis does not depend on $z$'s.

It is clear from Eq.(\ref{azxfin}) that the evaluation of
$F^6$ amplitudes is very different from the analogous $F^4$
computations.
The function $A(z,x)$, as well as the contributions of
other right-moving correlators, both ``split'' and ``non-split,''
involve non-trivial $z$-dependence.
Due to the presence of Cauchy kernels,
$A(z,x)$ is {\em not\/} symmetric under the permutations of $z$'s,
as contrasted
to the analogous correlators in the $F^4$
case where all (potentially)\footnote{Even without using supersymmetry,
the $t_8F^4$ structure follows
from the symmetry of right-moving correlators under
the permutations of vertex positions.
In fact, its two-loop coefficient is zero \cite{part1}.}
non-vanishing amplitudes are completely symmetric.
Thus the integration over the vertex positions is
highly non-trivial due to
the coupling of right-movers to the left-moving
Ka\v{c}-Moody correlators. The $F^6$ amplitudes do not seem to have
a topological origin.

\end{document}